\documentclass[aps,pra,twocolumn,groupedaddress,amsmath,amssymb]{revtex4-1}

\usepackage{graphicx}
\usepackage{dcolumn}
\usepackage{bm}
\usepackage{amsmath}
\usepackage{color}
\usepackage{multirow}
\usepackage{threeparttable}
\usepackage{float}

\def\apj{ApJ}                 
\def\apjl{ApJ}                
\def\aap{Astron. \&  Astroph.}                
\def\araa{ARA\&A}             
\def\apjs{ApJ}                
\def\apss{ApSS}		
\def\mnras{MNRAS}             
\def\pra{Phys. Rev. A} 			
\def\pre{Phys. Rev. E} 			

\newcommand{\bra}[1]{\langle #1|}
\newcommand{\ket}[1]{|#1\rangle}

\begin{document}


\title{A two-dimensional pseudospectral Hartree-Fock method for low-Z atoms in intense magnetic fields}


\author{Anand Thirumalai}
\email[SESE Exploration Postdoctoral Fellow, electronic address: ]{anand.thirumalai@asu,edu.}
\affiliation{School of Earth and Space Exploration, Arizona State University, Tempe, Arizona, USA, 85287}
\author{Jeremy S. Heyl}
\email[Canada Research Chair, electronic address: ]{heyl@phas.ubc.ca.}
\affiliation{University of British Columbia, Vancouver, British Columbia, V6T 1Z1}


\date{\today}

\begin{abstract}
The energy levels of the first few low-lying states of helium and lithium atoms in intense magnetic fields up to $\approx 10^8-10^9$~T are calculated in this study. A pseudospectral method is employed for the computational procedure. The methodology involves computing the eigenvalues and eigenvectors of the generalized two-dimensional Hartree-Fock partial differential equations for these two- and three-electron systems in a self-consistent manner. The method exploits the natural symmetries of the problem without assumptions of any basis functions for expressing the wave functions of the electrons or the commonly employed adiabatic approximation. It is seen that the results obtained here for a few of the most tightly bound states of each of the atoms, helium and lithium, are in good agreement with findings elsewhere. In this regard, we report new data for two new states of lithium that have not been studied thus far in the literature. It is also seen that the pseudospectral method employed here is considerably more economical, from a computational point of view, than previously employed methods such as a finite-element based approach. The key enabling advantage of the method described here is the short computational times which are on the order of seconds for obtaining accurate results for heliumlike systems.
\end{abstract}

\maketitle

\section{\label{sec:intro}Introduction}

The motivation to study atoms in magnetic fields of strength beyond the perturbative regime was in a large part due to the discovery of such fields being present in white dwarf stars \cite{Kemp1970, Angel1978, Angel1981} and neutron stars \cite{Trumper1977,Trumper1978}. The most commonly observed neutron stars - pulsars, have been observed to have magnetic fields on the order of $10^{7}$ - $10^{9}$T \cite{Ruder94}. Magnetars \cite{DT1992}, which are strongly magnetized neutron stars, can have magnetic field strengths well in excess of $10^{9}$T. White dwarf stars on the other hand have somewhat less extreme fields, albeit still high, $\sim10^{2}$ - $10^{4}$T \cite{Ruder94}. At such high field strengths, a Zeeman-type perturbative treatment of the field \cite{Landau} is not possible. The structure of atoms is considerably altered from the low field case. 

The problem of atoms in magnetic fields has been tackled by various researchers since the 1970's using a variety of different methods. In the literature, there exist numerous studies of hydrogen \cite{CK1972, Praddaude1972, SV1978, Friedrich1982, WR1980, RWRH1983, RWRH1984, Ivanov1988, VB2008, VB2002} and many recent studies of helium \cite{Proschel1982, Thurner1993, Ivanov1994, Jones1996, Jones1997, Jones1999, HH1998, MH2002, MH2007, Schmelcher2003, Schmelcher2003II, Schmelcher2002, Schmelcher2001, Schmelcher2000, Schmelcher1999, Schmelcher_helium2007, Ivanov1997, Ivanov2000, Wang2008} atoms in strong magnetic fields. There have also been studies conducted for molecules and chains of atoms for both hydrogen and helium atoms in strong to intense magnetic fields \cite{Lai1992, Lai1993, Lai1996, Medin2006I, Medin2006II, Schmelcher97hyd, Schmelcher2000hyd, Schmelcher2001hyd}. Moreover, our recent investigation \cite{TH2009} using single-configuration Hartree-Fock (HF) theory \cite{Hartree} was seen to yield accurate upper bounds for the binding energies of hydrogen and helium in strong magnetic fields. Our later study \cite{HT2010}, obtained accurate binding energies for helium and lithium atoms in strong magnetic fields using a pseudospectral method. This approach was seen to be computationally far more economical than using our earlier finite-element based approach \cite{TH2009}.

In sharp contrast to the somewhat simpler two-electron systems, there is very limited work available in the literature for atoms with more than two electrons. One of the first studies to investigate atoms in intense magnetic fields, in particular the iron atom, was by Flowers et al \cite{Flowers77} in 1977. This variational study extended the work due to the authors in Ref.~\cite{GK1975} and obtained binding energies of iron atoms and condensed matter in magnetic fields relevant to neutron stars. Errors in this study were later corrected by Muller \cite{Muller83}. Other methods included density functional studies \cite{PBJones85_1, PBJones85_2} and also employed the Thomas-Fermi-Dirac method \cite{Mueller71, Skjervold84} for estimating binding energies of atoms in intense magnetic fields. The first comprehensive HF studies of atoms with more than two electrons were carried out by Neuhauser et al \cite{Neuhauser86, Neuhauser87} for magnetic fields greater than $10^{8}$T, thus being directly relevant to neutron stars. Elsewhere, HF studies of atoms and molecules in intense magnetic fields were conducted by Demuer et al \cite{Demeur94}, with results consistent with previous findings. All of the above treatises, Refs.~\cite{Flowers77, GK1975, Muller83, PBJones85_1, PBJones85_2, Mueller71, Skjervold84, Neuhauser86, Neuhauser87, Demeur94}, concern themselves with magnetic fields in excess of $10^{8}$T, well into the so-called intense magnetic field regime. At these field strengths, the interaction of the electron with the nucleus of the atom becomes progressively less dominant, in comparison to its interaction with the field itself. One of the first studies to carry out a rigorous HF treatment of atoms with more than two electrons in strong or intermediate field strengths was Ref.~\cite{Jones1996}. Therein, they obtained estimates of the binding energies of a few low-lying states of lithium and carbon atoms, in low to strong magnetic fields. Elsewhere, Ivanov \cite{Ivanov1998} and Ivanov \& Schmelcher \cite{Ivanov1997, Schmelcher1999exch, Ivanov1999, Ivanov2000, Ivanov2001EPJD, Ivanov2001JPhB} have over recent years, carried out detailed HF and post-HF studies of atoms with more than two electrons using a numerical mesh-method for solving the unrestricted HF equations \cite{Ivanov1997}. The special meshes were so constructed as to facilitate finite-difference calculations in a two-dimensional domain using carefully selected mesh node points \cite{Ivanov2001}. They were able to ascertain the binding energies of the first few low-lying states of low-Z atoms such as lithium, beryllium and mid-Z atoms such as boron and carbon etc., using this method. Moreover, using a gaussian basis of functions for expressing the wave functions of the electrons \cite{Schmelcher2003, Schmelcher2003II, Schmelcher2002, Schmelcher2001, Schmelcher2000, Schmelcher1999, Schmelcher_helium2007}, adopting a full configuration-interaction method, Al-Hujaj \& Schmelcher \cite{Schmelcher_lithium2004, Schmelcher_beryllium2004} have been able to estimate the binding energies of lithium and beryllium atoms in strong or intermediate magnetic fields, thereby improving upon previously obtained results. The sodium atom in a strong magnetic field has also been studied by Gonzalez-Ferez \& Schmelcher \cite{Schmelcher_sodium2003} obtaining estimates for the binding energies. Elsewhere, low lying states of the lithium atom have also been studied in strong magnetic fields using a configuration-interaction method, employing the so-called freezing full-core method both with \cite{Qiao2000} and without \cite{Guan2001} correlation between electrons. Recently, Medin \& Lai \cite{Medin2006I, Medin2006II} have also studied atoms and molecules and infinite chains of condensed matter in magnetic fields greater than $10^{8}$T, using density-functional-theory. Mori et al \cite{MH2002, MH2007} have studied mid-Z atoms in strong to intense magnetic fields using perturbation theory as well, obtaining results consistent with previous findings. In recent years Engel and Wunner and co-workers  \cite{Engel2008, Engel2009, Schimeczek2013, Schimeczek2012, Klews2005} have computed accurate results for several atoms in magnetic fields relevant to neutron stars with a variety of techniques involving finite-element methods with B-splines both in the adiabatic approximation and beyond the adiabatic approximation with more than one Landau level. These highly accurate formulations employ a fast parallel Hartree-Fock-Roothan code, in which the electronic wave functions are solved for along the $z-$direction, with Landau orbitals (and combinations of more than one level in the latter studies) describing the remaining parts of the wave functions. Elsewhere, different \emph{ab initio} Quantum Monte-Carlo approaches \cite{Bucheler2007, Bucheler2008} have also been successfully employed for determining the ground states of atoms and ions in strong magnetic fields including lithium. Recently excited states of helium have also been computed quite accurately in intense magnetic fields using a fixed-phase Quantum Monte-Carlo approach \cite{Meyer2013}. The recent results of \citet[][]{Schimeczek2012, Schimeczek2013} for obtaining the ground state energies of atoms up to $Z=26$, were obtained within only a few seconds of computing time for helium and helium like atoms. Such speeds are essential for coupling atomic structure codes with atmosphere models and spectral analysis codes for magnetized white dwarfs and neutron stars. However, the work of \citet[][]{Schimeczek2013} only concerns itself with the ground state configurations. The primary aim of the current study is to provide an accurate method for investigating the ground and excited states of atoms, which can be computed in a matter of seconds on run-of-the-mill computer architectures. The method described here provides a speed-up of a factor of $10^4$ when compared with our earlier investigation using a finite-element discretization \cite[][]{TH2009}. Computational times are comparable to the recently developed quantum monte-carlo methods due to \citet[][]{Schimeczek2013}. The method described in this study can be run on a distributed architecture for investigating multi-electron atoms. 

It is also well known that post-HF methods such as configuration interaction (CI) or multiconfigration Hartree-Fock (MCHF) methods, yield considerable improvements with regard to the estimates of the upper bounds for the energies of various states. In the intermediate range of magnetic field strengths, where both the nucleus of the atom and the magnetic field have interactions with the electrons that are approximately equal in magnitude, the single configuration approximation then becomes increasingly ineffective with greater number of electrons. However, these methods are computationally more intensive than a single configuration calculation. Thus far, the most accurate CI methods involve decomposing the wave functions into a Gaussian basis set relying upon separation of variables in cylindrical coordinates \cite[e.g.][]{Schmelcher_lithium2004, Schmelcher_beryllium2004}. These methods do however require a large set of basis functions. On the other hand, MCHF methods would require fewer basis functions, as the orbitals get optimized during the computation with the coefficients \cite[e.g.][]{CFF1997}. Separation of variables and/or basis decompositions speed up the computation in these post-HF methods considerably. However, there do not exist hitherto, any fully two-dimensional (2D) post-HF studies of multi-electron atoms in intense magnetic fields. This is partly due to the computational overhead associated with adopting a fully 2D picture. Central to the development of such a method would be the fast and accurate computation of the single-configuration problem in a full 2D framework without any basis expansions and separation of variables. Wave functions so determined could be used directly in 2D configuration-interaction calculations or the problem could be cast into a MCHF framework. Moreover, obtaining accurate estimates of the energy levels of atoms, in particular low-Z atoms, in strong and intense magnetic fields will ultimately facilitate a proper understanding of the spectra of neutron stars and white dwarf stars. In order to achieve this, a large amount of accurate data needs to be made available for not only binding energies of several different states, but also oscillator strengths, as well as estimates of the energies associated with both bound-free and free-free electron transitions in a variety of magnetic field strengths. There is also the added complication of strong electric fields affecting the spectra of atoms moving perpendicular to the magnetic field; however that being said, estimating the binding energies due to intense magnetic fields is part of that picture. Thus the central aim of the current work is to take a step in that direction and provide accurate and readily calculable estimates of the binding energies of the first few low-lying states of the simplest low-Z atoms; helium and lithium, in strong and intense magnetic fields using a two dimensional single-configuration pseudospectral method of solution. 

The method outlined in the current study is an extension of the method developed in our two previous studies Refs.~\cite{TH2009,HT2010}.

\section{\label{sec:HF}The HF Equations}

We shall begin with the generalised single-configuration HF equations for an atom with $n_e$-electrons and nuclear charge $Z$, in a magnetic field that is oriented along the $z$-direction. A derivation of the single-configuration HF equations can be found in our earlier work \cite{TH2009}, here we shall present only the salient points. The single configuration HF equation can be written in cylindrical coordinates as, where the length scale is in units of Bohr radii and the energy is scaled in units of Rydberg energy in the Coulomb potential of charge $Ze$ (see below for definitions).
\begin{widetext}
\begin{eqnarray}
\left[-\nabla^{2}_{i}(\rho_{i},z_{i})+\frac{m_{i}^{2}}{\rho_{i}^{2}}+2\beta_{Z}(m_{i}-1)
+\beta_{Z}^{2}\rho_{i}^{2}
-\frac{2}{r_i}\right]\psi_{i}\left(\rho_{i},z_{i}\right)
+\frac{2}{Z}\sum_{j\neq
  i}\left[\Phi_{D}\psi_{i}(\rho_{i},z_{i})-\alpha_{E}\psi_{j}(\rho_{i},z_{i})\right]=\epsilon_{i}\psi_{i}\left(\rho_{i},z_{i}\right),\label{eq:1}
\end{eqnarray}
where \begin{math}i,j=1,2,3,...,n_e\end{math} and
\begin{math}r_i=\sqrt{\rho_i^2+z_i^2}.\end{math}  Please note that the three-dimensional momentum operator has been split into two parts; $\nabla_{i}^2(\rho_i,z_i)$ which are the $\rho-$ and $z-$ parts of the Laplacian and $m_i^2/\rho_i^2$ which is the azimuthal part. The total Hartree-Fock energy of
the state is given by
\begin{equation}
\varepsilon_{total}=\sum_{i}\epsilon_{i}-\frac{1}{2}\frac{2}{Z}\sum_{j\neq i}\left[\bra{\psi_{i}(\rho_{i},z_{i})}\Phi_{D}\ket{\psi_{i}(\rho_{i},z_{i})}-\bra{\psi_{i}(\rho_{i},z_{i})}\alpha_{E}\ket{\psi_{j}(\rho_{i},z_{i})}\right].\label{eq:2}
\end{equation}
The direct ($\Phi_D$) and exchange ($\alpha_E$) interactions are determined according to the method outlined in Ref.~\cite{TH2009}, as the solutions of the elliptic partial differential equations for the potentials given by,
\begin{equation}
\nabla^{2}_{i}\Phi_{D}=-4\pi|\psi_{j}(\rho_{i},z_{i})|^{2}
\label{eq:3}
\end{equation}
and
\begin{equation}
\left[
\frac{1}{\rho_{i}}\frac{\partial}{\partial\rho_{i}}
	\left(\rho_{i}\frac{\partial}{\partial\rho_{i}}\right)-\frac{(m_{i}-m_{j})^{2}}{\rho_{i}^{2}}+\frac{\partial^{2}}{\partial z_{i}^{2}}\right]\alpha_{E}(\rho_{i},z_{i})
=-4\pi\psi_{j}^{*}(\rho_{i},z_{i})\psi_{i}(\rho_{i},z_{i})
\label{eq:4}
\end{equation}
\end{widetext}
where $\psi_i$ and $\psi_j$ are the wave functions of the $i^\textrm{th}$ and $j^\textrm{th}$ electrons. The wave function of a given configuration of electronic orbitals is assumed to be given by a single Slater determinant as,
\begin{equation}
\Phi=A_{n_e}\left(\tilde{\psi}_{1}, \tilde{\psi}_{2}, \tilde{\psi}_{3}, ..., \tilde{\psi}_{n_e-1},\tilde{\psi}_{n_e}\right),
\label{eq:5}
\end{equation}
where \begin{math}A_{n_e}\end{math} is the anti-symmetrization operator. The individual electronic wave functions $\tilde{\psi}_{i}$ are given by,
\begin{equation}
\tilde{\psi}_{i}= \psi_{i}(\rho_{i},z_{i})e^{im\phi_{i}}\chi_{i}(s_{i}),
\label{eq:6}
\end{equation}
where \begin{math}i\end{math} labels each of the $n_e$ electrons. The two-dimensional single particle wave functions \begin{math}\psi_{i}(\rho_{i},z_{i})\end{math} are taken to be real functions.

Integration with respect to the azimuthal coordinate, $\phi$, has been carried out, prior to writing the result in Eq.~(\ref{eq:1}) above. The contribution due to electron spin has also been averaged out \textit{a priori}. It is to be mentioned in this regard, that in the current study we shall only be concerned with fully spin-polarised states (FSP), in other words all the electrons of the atom are assumed to be anti-aligned with the magnetic field. Such states have an exchange interaction between the electrons providing an extra coupling term in the HF equations, $\alpha_E$. Additionally, FSP states are seen to be the most tightly bound states in the intense field regime. The extension to partially spin-polarised configurations is easily achieved by eliminating the exchange term in the HF equations. In the current study, we have chosen to work in units of Bohr radii along with the definitions given below. 

The Bohr radius for an atom of nuclear charge $Z$ is given by $a_{B}/Z$, where $a_B=\hbar/\alpha m_e c$ is the Bohr radius of the hydrogen atom. The magnetic field strength parameter $\beta_Z$, is given by the expression $\beta_Z=B/(Z^2B_0)$, where $B_0$ is the critical field strength at which point the transition to the intense magnetic field regime occurs \cite{Ruder94}. This is defined as  $B_{0}=(2\alpha^{2}{m_{e}}^{2}c^{2})/(e\hbar) \approx 4.70108 \times 10^5$T. Thus, beyond a value of $\beta_Z\approx1$, the interaction of the electron with the nucleus becomes progressively less dominant, as $\beta_Z$ increases. Based upon the above definition of $\beta_Z$, it is convenient to classify the field strength \cite{Jones96} as low ($ \beta_Z \leq 10^{-3}$), intermediate, also called strong ($ 10^{-3} \leq \beta_Z \leq 1 $) and intense or high ($  1 \leq \beta_Z \leq \infty $). These definitions of the different magnetic field strength regimes are useful to remember when discussing the results in the latter part of this paper and for distinguishing between ``strong" and ``intense" magnetic field strengths. The energy parameter of the $i^{th}$ electron is defined as $\epsilon_{i}=E_{i}/(Z^{2}E_{\infty})$, with $E_{\infty}=\frac{1}{2}\alpha^2m_e c^2$, the Rydberg energy of the hydrogen atom. For brevity we shall refer to the units of energy as $E_{Z,\infty}$, which should be remembered as the Rydberg energy in the Coulomb potential of charge $Ze$. The quantity $\alpha=e^2/(4\pi\epsilon_{0}\hbar c)\approx1/137$ is the fine structure constant. In the current study, all the physical constants were used in SI units. Additionally, the magnetic field $B$, is taken to be in units of Tesla. Eq.~(\ref{eq:1}) represents the $N$-coupled Hartree-Fock equations in partial differential form for an $N$-electron system with nuclear charge $Z$. The equations are coupled through the exchange interaction term between the electrons and as such the system of equations is solved iteratively. The discussion that follows is arranged in the following manner. In Section~\ref{sec:numer}, we shall describe the numerical methodology employed in the current study. For solving the system of partial differential equations we adopted a pseudospectral approach. The interested reader will find the entire method of discretizing and setting up the discrete problem using this approach in the appendix. Readers familiar with the method may wish to only read Sections~\ref{sec:numer}-\ref{sec:conclusion}. In Section~\ref{sec:results} the results of the current study are provided alongside a discussion. Section~\ref{sec:conclusion} contains the the conclusions of the current study and avenues for further work are also discussed therein.

\section{\label{sec:numer}Numerical Details}

The numerical solution of the coupled eigenvalue problem in Eq.~(\ref{eq:1}) proceeds via the so-called self consistent field (SCF) method due to Hartree \cite{Hartree}. First we find a solution to the hydrogenic problem, i.e., Eq.~(\ref{eq:1}) without the direct and exchange interactions. This yields ionic single electron hydrogenic wave functions in the Coulomb potential of charge $Ze$ forming the initial estimates for the HF iterations. Second, using these estimates the elliptic partial differential equations for the direct and exchange interaction potentials in Eqs.~(\ref{eq:3}) and (\ref{eq:4}) are solved. With these potentials now obtained, the coupled HF problem including the direct and exchange interactions in Eq.~(\ref{eq:1}), is solved as an eigensystem. The exchange interactions which couple the equations are expressed using wave-functions from the previous iteration to solve the eigenvalue problem for each electron \cite{Slater1951}. The eigenvalues obtained are the individual particle energies $\epsilon_i$ and the normalized eigenvectors are the wave functions, $\psi_i(\rho_i,z_i)$. The SCF iterations then proceed with the updated electron wave functions and the steps from the second step described above, are repeated until convergence.

\subsection{\label{sec:domain}Domain Discretization}

The very first step in this direction is the discretization of the physical domain of the problem. By virtue of azimuthal symmetry and parity with respect to the $z=0$ plane, it is sufficient to restrict the domain of the problem to $ 0  \leqslant \rho, z \leqslant \infty$ \cite{Ruder94, TH2009}. 
However, we solve the problem not in this semi-infinite domain, but rather in a finite but sufficiently large domain of size $\rho_{\textrm{max}} \times z_{\textrm{max}}$. Whenever the domain of a problem is restricted, this introduces an error since the computational problem then becomes an approximation of the actual problem. Therefore in order to eliminate this error associated with truncating the domain, we employ a sequence of domains of increasing sizes, obtaining a converged result in the limit of the computational domain approaching the size of the physical domain of the problem.


In our computations, the size of the computational domain $\rho_{\textrm{max}}$ and $z_{\textrm{max}}$ (in units of Bohr radii) are given by,
\begin{equation}
\rho_{\textrm{max}} ~,~ z_{\textrm{max}} = \frac{100 \eta}{1+\textrm{log}_{10}(\beta_Z)},
\label{eq:compactification}
\end{equation} 
where $\eta=1/4, 1/2, 1, 2$ is a scaling factor used for setting up computations in a sequence of increasing domain sizes. The effect of the logarithmic term $\textrm{log}_{10}(\beta_Z)$, in the denominator is that it naturally makes the domain larger or smaller, depending on whether $\beta_Z < 1$ or $\beta_Z > 1$, respectively. With the maximum domain size thus defined, we can then compactify the finite domain $[0,\rho_{\textrm{max}}] \otimes [0,z_{\textrm{max}}]$ to $[-1,1] \otimes [-1,1]$ with the transformation,
\begin{equation}
x = \textrm{log}_{10} (1 + \rho \alpha_{\rho} ) - 1
\label{eq:transform_rho}
\end{equation}
and
\begin{equation}
y = \textrm{log}_{10} (1 + z \alpha_{z} ) - 1,
\label{eq:transform_z}
\end{equation}
where $\alpha_{\rho}=99/\rho_{\textrm{max}}$ and $\alpha_{z}=99/z_{\textrm{max}}$. Note that in our calculations we employed a square domain for achieving the best possible internally consistent convergence. Therefore in our work $\rho_{\textrm{max}} = z_{\textrm{max}}$ and therefore $\alpha_\rho = \alpha_z \equiv \alpha$, but the possibility remains for using different sizes and scalings in the two orthogonal directions for optimizing computational effort particularly in the intense field regime.

With this compactification scheme, it is then possible to employ a Chebyshev-Lobatto spectral collocation method \cite{Trefethen} with discrete points on the domain $[-1,1] \otimes [-1,1]$. The details regarding the pseudospectral approach developed for solving the HF equations in Eq.~(\ref{eq:1}) can be found in the appendix. An atomic structure software package based on this method was written in the high level programming language MATLAB$^{\textregistered}$ making particular use of its fast matrix manipulation algorithms. 
The eigenvalue problems described in Eqs.~(\ref{eq:52}) and (\ref{eq:64})
are solved by discretizing the equations and solving
the resultant algebraic eigenvalue problem. This ultimately produces a sparse matrix for the coupled eigenvalue problem (see the appendix and Fig.~(\ref{fig:fig5}) therein). Thus we can take advantage of this fact and employ a sparse matrix generalized eigensystem solver; the widely used package ARPACK which utilizes the implicitly restarted Arnoldi method (IRAM)
\cite{Arnoldi1951,Sorensen1992,ARPACK}. The key advantage of employing
IRAM is that the memory storage requirements are considerably reduced since the Arnoldi factorization generates a small Krylov subspace using a few basis vectors and only a handful of eigenvalues are computed in a given portion of the spectrum \cite{ARPACK, Saad1984, Sorensen1992}. Since the Hamiltonian matrix that we are solving only has a few bound state solutions, employing IRAM for computing only a portion of the spectrum is therefore highly desirable and saves considerable computational effort, particularly for a two-dimensional problem where the discretisation can generate large matrices.
%
%
This method was found to yield accurate
results for the energy eigenvalues of the first few eigenstates of
helium and lithium in intense magnetic fields. It was seen that generating a
Krylov subspace with about $50$ to $250$ basis vectors was sufficient for
determining around $15$ to $100$ eigenvalues in the vicinity of a given shift
($\sigma$), by employing the shift-invert algorithm
\cite{ARPACK}. Runs were carried out for different values of the
magnetic field strength parameter \begin{math} \beta_Z \end{math}, in
the range \begin{math} 5 \times 10^{-1} \leq \beta_Z \leq
  10^3 \end{math}, for the cylindrical pseudospectral code. A typical
tolerance of around $10^{-10}$ was employed for the internal errors of
ARPACK. It was observed during our runs that fast convergence was achieved; within about $3-6$ HF iterations. A convergence criterion for the HF iterations was employed wherein the difference between the HF energies for two consecutive iterations was tested. Typically, a tolerance on the order of $10^{-6} E_{Z,\infty}$ was employed. Once the HF iterations attained convergence for a given level of mesh refinement, the total energy of the Hartree-Fock state under consideration is reported according to Eq.~\ref{eq:2}. Additionally for testing convergence of the pseudospectral
method, we employed up to six different levels of mesh refinement for lithium and seven for helium,
ranging from coarse to fine mesh i.e., $N=21,31,41,51,61,71~\textrm{and}~ 81$ mesh points in each direction, for a given domain size, $\rho_{\textrm{max}} \times z_{\textrm{max}} $. Once a converged solution was obtained for a given domain size, using a sequence of meshes, a different value for the scaling parameter $\eta$ was employed and the calculation was carried out again.

\subsection{\label{sec:speed}Computational speeds}
\begin{figure}[h]
\begin{center}
\includegraphics[width=3.5in, scale=1.0]{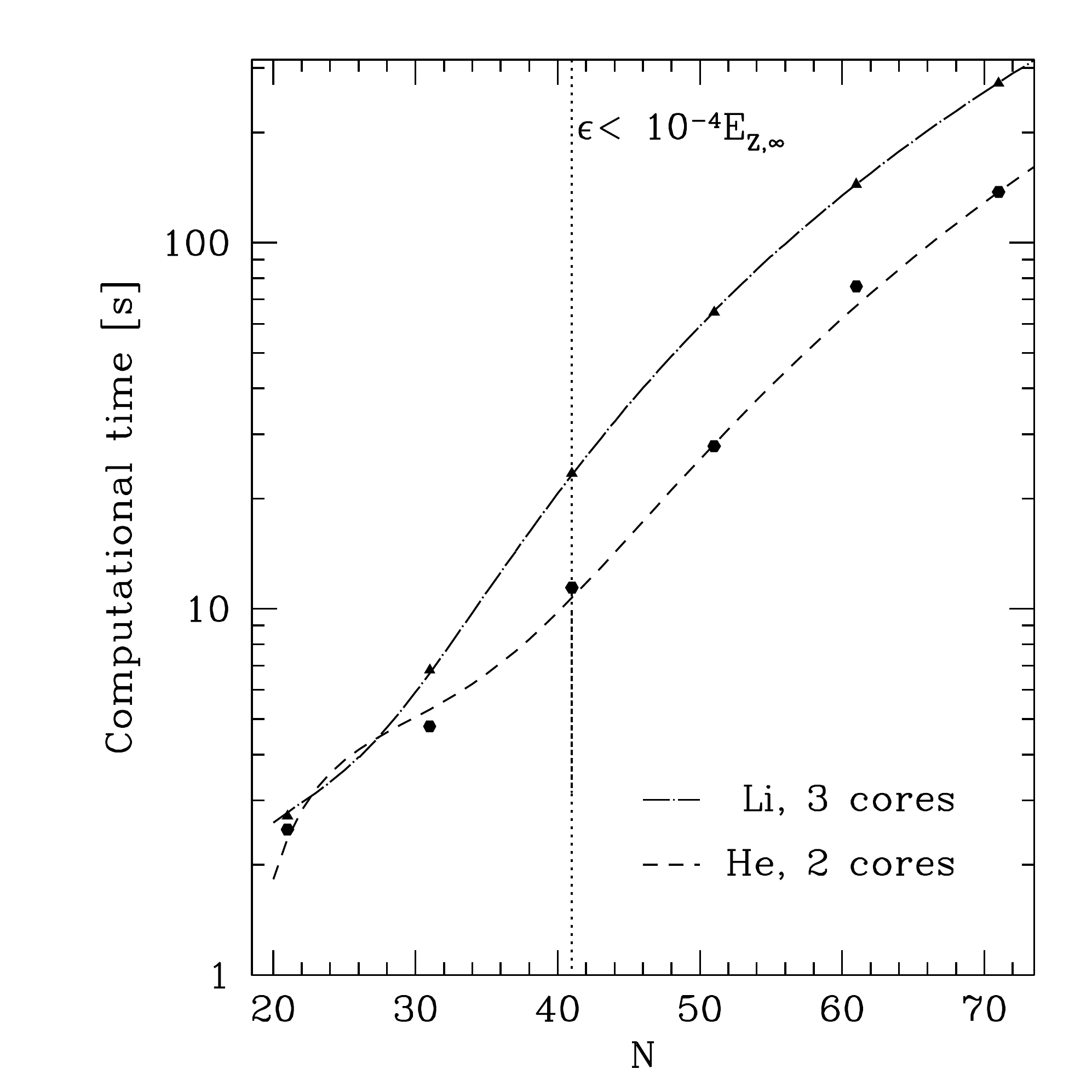}
\end{center}
\caption{Computational times shown as a function for the number of grid points, for typical calculations for both helium and lithium. The dotted vertical line corresponds to the number of grid points typically required to attain an accuracy of $\epsilon \approx 0.1\% \equiv 10^{-4}E_{Z,\infty}$. The number of cores or processors employed by the parallel code is equal to the number of electrons in the atom. The dependence of the computational time on the number of grid points is roughly $\mathcal{O}(N^3)$, while the dependence on the number of electrons is more or less linear, i.e., $\mathcal{O}(n_e)$.}
\label{fig:compute_times}
\end{figure}

One of the key enabling advantages of the method adopted in this study is the reduced computational time. The gain in speed is largely due to two reasons. First, since the eigenvalue problem represents a smooth elliptic one whose solution does not have discontinuities, a pseudospectral approach yields accurate results with the use of a small number of discretisation points or grid elements. Second, our code is implemented on a parallel architecture, where the number of processors (or cores) employed is equal to the number of electrons in the atom. With this trivial parallelization, the eigenvalue computation then proceeds in parallel for all the electrons, and the computational time is then determined by the electron for which the eigenvalue computation takes the longest time.
The calculations for lithium for example took a total of about $20$ seconds of computing time running on three Intel$^{\textregistered}$ Xeon$^{\textregistered}$ E5620 $2.4$~GHz processors, for obtaining accuracy $\stackrel{_<}{_\sim}10^{-4}E_{Z,\infty}$. The fastest calculations we observed were for helium where the calculations converged on the order of about $10$ seconds, with computations carried out on two processors, for achieving the same level of accuracy. These speeds are comparable to the recent results of \citet[][]{Schimeczek2013} who in their calculations achieve considerable speed up by first pre-calculating the inter-electron potentials and second by utilizing Landau states to describe the wave functions in the $\{\rho,\phi\}$ directions, thus only having to solve for the unknown $z$-part of the wave functions. In our approach, we are not restricted to the adiabatic approximation and the unknown wave function that is solved for is a two-dimensional one, and the eigenvalue problem is therefore a fully coupled two-dimensional problem. Additionally, the inter-electron potentials in our approach are recalculated as the wave function gets optimized in each HF iteration, by solving elliptic partial differential equations, which adds to the computational times. This combination of speed and accuracy achieved in our computations is a result of spectral convergence. Figure~\ref{fig:compute_times} shows the dependence of computational time on both the level of domain discretization and the number of electrons in the problem, which is equal to the number of processors employed. The computational time increases roughly as $\mathcal{O}(N^3)$, where $N$ is the number of grid points in each direction. The lines drawn through the data in Fig.~\ref{fig:compute_times} correspond to polynomials of degree $3$. 
It can also be seen that the compute times increase with the number of electrons ($n_e$). However, it can be seen that this dependence scales more or less linearly, i.e., $\mathcal{O}(n_e)$. As a result, we expect that the method can be extended to mid-Z atoms with $n_e>3$, without incurring a large computational overhead.
%
%
%
\subsection{\label{sec:limitations}Limitations of the approach}
The atomic structure package developed in this study also has certain limitations. First, since the method employed presently requires a finite domain, computations are required to be carried out on a sequence of domains to eliminate errors due to domain truncation. This could be circumvented by employing an adaptive scheme, wherein the wave functions at the exterior domain points ($\rho,z \rightarrow \infty$) are required to fall below a certain threshold or tolerance \citep[e.g.][]{Schimeczek2013} while employing an exponential grid. This would however complicate the pseudospectral approach, although a key feature of using Chebyshev-Lobatto points is that the density of points is greatest at the end points of the domain, which may aid in resolving the wave function at the outer limits of the domain, should such a scheme be implemented.
\textbf{Second,} the current work does not include relativistic corrections to the energies. For the magnetic field strengths considered herein, the relativistic corrections to the energies were estimated using the scaling formula in Ref~\cite{Poszwa2004}. Their results for the hydrogen atom were used for this purpose and the corrections were estimated to be on the order of $10^{-6} E_{Z,\infty}$. This was seen to be smaller than the numerical errors arising from convergence of the entire numerical method including the extrapolation to the limit of a semi-infinite domain. Thus, while relativistic corrections are important, it was not possible to account for them accurately in the current study.
Presently, the results are presented alongside a discussion in the following section.

\section{\label{sec:results} Results and Discussion}

The atomic structure software package developed in the current study extends our earlier computations for atoms in strong magnetic fields \cite{TH2009,HT2010}, towards the intense field regime, $\beta_Z \gg 1$. The states that were considered in this study are labelled using both the field-free and strong-field notations for the sake of clarity; these can be found in Table~\ref{tab:table1} which lists these different states of helium and lithium. In the presence of a magnetic field states can be characterized using the notation $\nu^{2S+1}\textrm{M}^{\pi_z}$, where $\textrm{M}=\Sigma_i m_i$ is the total $z-$ component of angular momentum. The summation is over all the electrons in the atom. This then forms a manifold within which different sub-spaces exist. The quantum number $\nu$ counts the excitation level within a given $\textrm{M}-$manifold and sub-space symmetry. The spin multiplicity is given in the usual way as $2S+1$. Finally, the $z-$parity of the state is indicated using $\pi_z= \pm 1$, indicating positive or negative parity. We studied the six most tightly bound states of each of these two atoms in the intense magnetic field regime ($\beta_Z \gg 1$). Within a given parity sub-space, typically there are crossovers that occur as the magnetic field is reduced and the reader is referred to Ref.~\cite{Ivanov2000} for an excellent discussion regarding ground state crossovers. Within each given parity sub-space, we considered the most tightly bound state in the intense field regime. 
\begin{table}[h]
\centering
\caption{The different states of helium and lithium considered in this study, listed using both intense-field and field-free notation.}
\begin{tabular}{c@{\hspace{3mm}}c@{\hspace{3mm}}c@{\hspace{5mm}}}
\hline
\hline
& Intense-field ~ & ~ Field-free \\
\hline
\multirow{6}{*}{Helium} ~~ ~~ & $1^3(-1)^+$ & $1s_02p_{-1}$ \\
& $1^3(-1)^-$ & $1s_03d_{-1}$ \\
& $1^3(-2)^+$ & $1s_03d_{-2}$ \\
& $1^3(-2)^-$ & $1s_04f_{-2}$ \\
& $1^3(0)^+$ & $1s_02s_{0}$ \\
& $1^3(0)^-$ & $1s_02p_{0}$ \\
\hline
\multirow{6}{*}{Lithium} ~~ ~~ & $1^4(-3)^+$ & $1s_02p_{-1}3d_{-2}$ \\
& $1^4(-3)^-$ & $1s_02p_{-1}4f_{-2}$ \\
& $1^4(-2)^+$ & $1s_02s_03d_{-2}$ \\
& $1^4(-2)^-$ & $1s_02p_{-1}3d_{-1}$ \\
& $1^4(-1)^+$ & $1s_02s_{0}2p_{-1}$ \\
& $1^4(-1)^-$ & $1s_02p_{0}2p_{-1}$ \\
\hline
\hline
\end{tabular}
\label{tab:table1}
\end{table}

\subsection{\label{He_results} The Helium Atom}

\begin{figure}[h]
\begin{center}
\includegraphics[width=3.5in, scale=1.0]{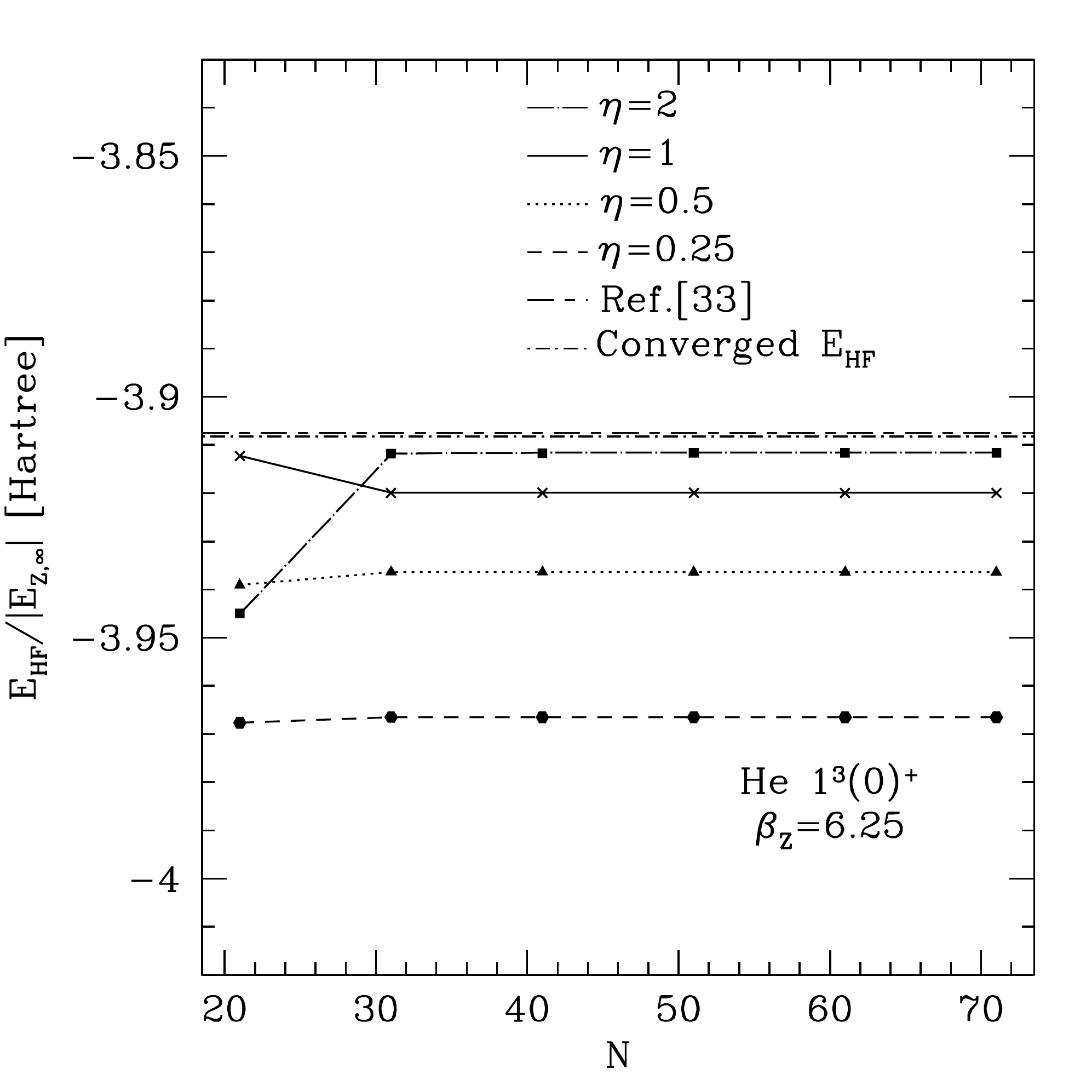}
\end{center}
\caption{Convergence of the binding energy with mesh refinement for four different domain sizes for the $1^3(0)^+$ state of Helium. The level of mesh refinement corresponds to $N$ points in each of the two orthogonal directions. The size of matrix of the coupled eigenvalue problem for a given level of mesh refinement is given by  $n_e (N-1)^2 ~ \times ~ n_e (N-1)^2$, where $n_e$, is the number of electrons, which for helium is two. The levels of mesh refinement employed correspond to
  $N=21,31,41,51,61$ and $71$ points in each of the $x-$ and $y-$. The exponential convergence of the spectral method can readily be seen. }
\label{fig:1s_2s_convergence}
\end{figure}

\begin{figure}[h]
\begin{center}
\includegraphics[width=3.5in, scale=1.0]{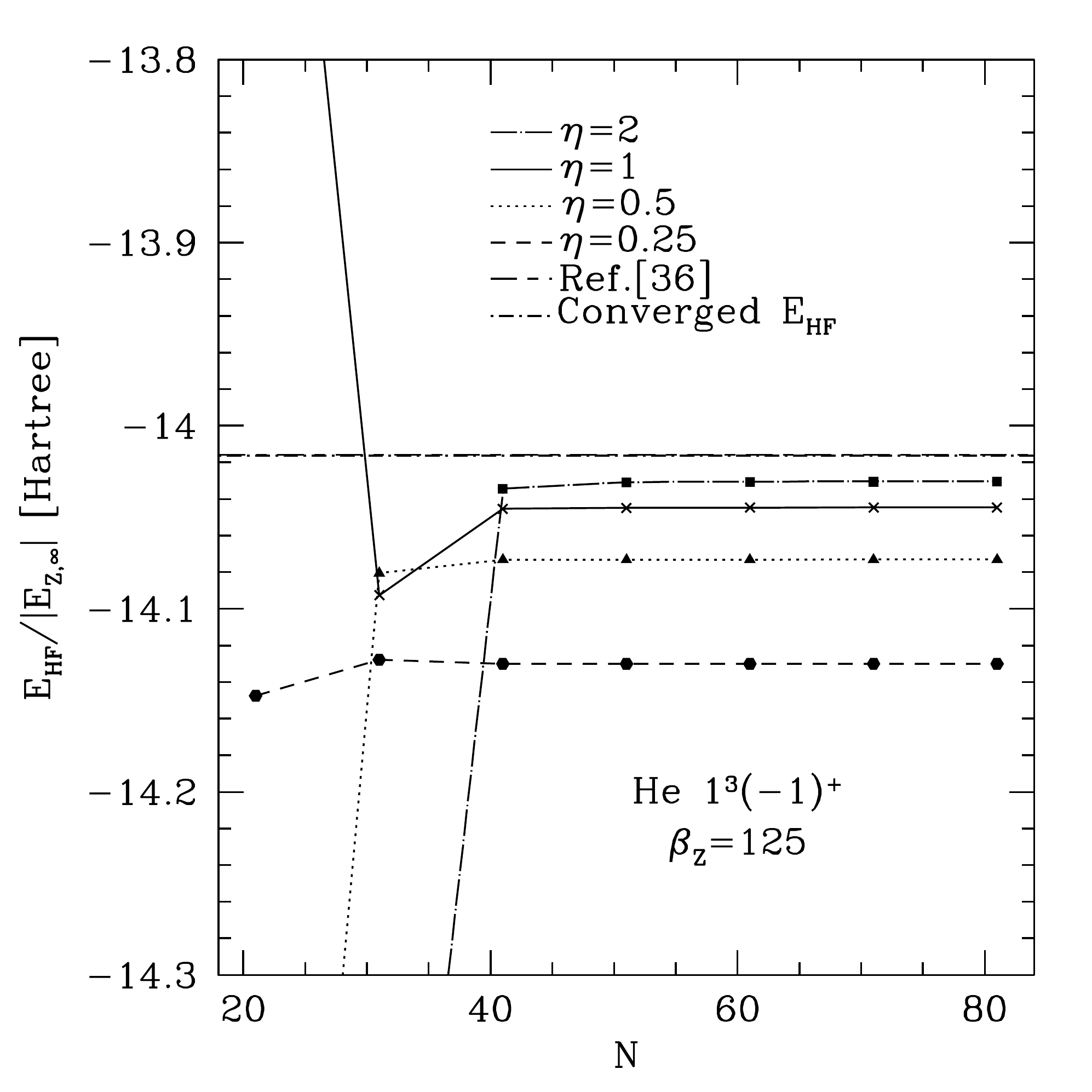}
\end{center}
\caption{Convergence of the binding energy with mesh refinement for four different domain sizes for the $1^3(-1)^+$ state of Helium. The level of mesh refinement corresponds to $N$ points in each of the two orthogonal directions. The size of matrix of the coupled eigenvalue problem for a given level of mesh refinement is given by  $n_e (N-1)^2 ~ \times ~ n_e (N-1)^2$, where $n_e$, is the number of electrons, which for helium is two. The levels of mesh refinement employed correspond to
  $N=21,31,41,51,61, 71$ and $81$ points in each of the $x-$ and $y-$ directions for helium. The exponential convergence of the spectral method can readily be seen for computations in the different domain sizes. }
\label{fig:1s_2p_1_convergence}
\end{figure}

For the states of helium listed in Table~\ref{tab:table1}, eigenvalues were determined using the numerical method described in Section~\ref{sec:numer}. We began with the lowest value of the domain scaling parameter $\eta=1/4$. This yielded a domain with dimensions given according to Eq.~(\ref{eq:compactification}), and this domain size depends on $\beta_Z$. HF energies were then calculated using up to six different levels of mesh refinement in the domain. This enabled us to extrapolate the results to the limit of infinitely fine mesh, for a given domain size. Thereafter, the domain was rescaled to larger and larger values, corresponding to $\eta=1/2, 1, 2$ and the computations repeated once again with different levels of mesh refinement for each value of $\eta$ so chosen. Then, using the extrapolated values of the HF energy corresponding to infinitely fine mesh for each of the four domain sizes, a subsequent extrapolated value of the the HF energy ($E_{HF}$) was obtained, in the limit of the domain size approaching infinity. These are then the ``converged $E_{HF}$" values reported in Figs.~\ref{fig:1s_2s_convergence} and ~\ref{fig:1s_2p_1_convergence} as well as in Table~\ref{tab:table2}.  These figures show typical examples of \emph{spectral convergence} that was achieved using our pseudospectral implementation, wherein the internal errors of the procedure diminish in an exponential fashion with mesh refinement.

Figure~\ref{fig:1s_2s_convergence} shows HF energies obtained from calculations for the $1^3(0)^+$ state of helium, in a magnetic field of strength $\beta_Z=6.25$. Meanwhile Fig.~\ref{fig:1s_2p_1_convergence} shows convergence data for the $1^3(-1)^+$ state of helium in a magnetic field corresponding to $\beta_Z=125$. The spectral method converged to better than typically $ < 10^{-6} E_{Z,\infty}$ with mesh refinement within each domain size. At the higher end of the intense field regime, $\beta_Z \stackrel{_>}{_\sim} 500$ there was some loss of accuracy in the convergence. In this region, the spectral method only converged to $\approx 10^{-4} E_{Z,\infty}$ with mesh refinement, particularly for the largest domain scaling parameter employed; $\eta=2$. These HF energies for different mesh refinements were extrapolated to the limit of infinitely fine mesh using an exponential function which typically took the form $ae^{bx}+ce^{dx}$. The errors associated with the extrapolation procedure were typically on the order of $10^{-6} E_{Z,\infty}$ with a normalized $R-$squared value typically $>0.999$ for the interpolating function employed. Again, at the upper end of the intense magnetic field regime, we noticed some loss of accuracy as the states become tightly bound, and for $\beta_Z \stackrel{_>}{_\sim} 500$ the extrapolation procedure had an error on the order of $10^{-5} E_{Z,\infty}$ with a normalized $R-$squared of $\approx 0.98$ on average. For the extrapolation to infinitely fine mesh, the average area per unit grid size in the domain ($A_E \approx \rho_{\textrm{max}} z_{\textrm{max}}/N^2$), was taken as the independent variable and the energies extrapolated to the limit of $A_E \rightarrow 0$, corresponding to infinitely fine mesh. 

It can also be seen in Figs.~\ref{fig:1s_2s_convergence} and \ref{fig:1s_2p_1_convergence} that by increasing the domain size (by increasing the domain scaling parameter $\eta$), the binding energies decrease. This clearly shows that a confinement energy is introduced when the domain of the problem is truncated which can lead to an overestimation of the binding energy. Additionally, in Figs.~\ref{fig:1s_2s_convergence} and \ref{fig:1s_2p_1_convergence}, it is readily seen that when the area of the domain is doubled, the difference in the binding energies are roughly halved, with each doubling. For example, $(E_{HF}^{\eta=1/2}-E_{HF}^{\eta=1/4}) \approx 2 (E_{HF}^{\eta=1}-E_{HF}^{\eta=1/2}) \approx 4 (E_{HF}^{\eta=2}-E_{HF}^{\eta=1})$, illustrating quadratic convergence of the method with increasing domain size. Graphically this can be seen in the approximate halving of the distance between the horizontal lines joining the data points, when the domain area is doubled. This convergence can be expressed as,
\begin{equation}
\begin{split}
E_{HF} &\approx E_{HF}^{\eta=1/4} + \Delta^{1}\left(1+ \frac{1}{2} + \frac{1}{4} + ... + \infty  \right) \\
&\approx E_{HF}^{\eta=1/4} + 2 \Delta^{1},
\label{eq:convergence}
\end{split}
\end{equation}
where $\Delta^{1}=E_{HF}^{\eta=1/2}-E_{HF}^{\eta=1/4}$. This can be readily seen in Figs.~\ref{fig:1s_2s_convergence} and \ref{fig:1s_2p_1_convergence}, where the converged $E_{HF}$ value is approximately twice as displaced from the data corresponding to $E_{HF}^{\eta=1/4}$ in comparison to the data corresponding to $E_{HF}^{\eta=1/2}$. In order to numerically obtain a converged HF energy in the limit of the domain size approaching infinity, we employed an extrapolating function of the form $ax^{1/2}+b$. The ordinates in this case were the four different converged HF energies in the limit of infinitely fine mesh in each of the four different domains, and the abscissae were the inverse domain areas, i.e. $(\rho_{\textrm{max}} z_{\textrm{max}})^{-1}$. Thus, extrapolating to zero inverse area corresponding to an infinite domain size yields the final converged HF energy indicated in Figs.~\ref{fig:1s_2s_convergence} and \ref{fig:1s_2p_1_convergence}, and it is these extrapolated values that are reported in Tables~\ref{tab:table2} and \ref{tab:table3}. The error in the extrapolation to the limit of an infinite domain size was on the order of $10^{-6} E_{Z,\infty}$ with a normalized $R-$squared value of $>0.999$ for the interpolating quadratic function. This extrapolation to the limit of infinite domain size eliminates the effect of introducing a confinement energy when the domain of the problem is truncated.

Inspection of Figs.~\ref{fig:1s_2s_convergence} and \ref{fig:1s_2p_1_convergence} reveals that the single-configuration converged HF energy in the limit of an infinite domain size is in good agreement with the corresponding results due to Ref.~\cite{Schmelcher1999} and Ref.~\cite{Schmelcher2001}. The results obtained in this study therefore represent an upper bound for \emph{uncorrelated} HF energies. These estimates can be improved using correlated post-HF methods, such as CI or MCHF methods, and the $2-$D wave functions computed by our method can form the initial inputs. The difference between our results and those of Ref.~\cite{Schmelcher1999} or Ref.~\cite{Schmelcher2001} therefore convey a sense of how large the effect of electron correlation can be even for helium. This effect typically increases with the addition of more electrons, becoming increasingly more pronounced, and therefore it becomes important that for accurate binding energies, electron correlation be taken into account. However, this is outside the scope of the current single-configuration calculation and we leave such a full $2-$D post-HF computation for a future undertaking.

Moreover, by comparing Figs.~\ref{fig:1s_2s_convergence} and \ref{fig:1s_2p_1_convergence} it can be seen that convergence with regards to mesh refinement is much cleaner when the magnetic field strength is lower. In Fig.~\ref{fig:1s_2s_convergence} it can be seen that regardless of the domain scaling parameter $\eta$, a converged solution (to well within $1\%$) is obtained with as few as about $31-41$ points whereas, for larger $\beta_Z$, the number of points required is typically greater, $41-51$, for obtaining a similarly converged result. This is due to the fact that the geometry of the wave functions becomes increasingly extreme with increasing magnetic field strength and a greater number of discretization points are therefore required to accurately resolve the wave functions. Overall, we observed that on average about $41-51$ discretisation points were enough for achieving convergence to well within $1\%$.


The converged HF energies for the positive parity states of helium are given in Table~\ref{tab:table2}, while those for the negative parity states are shown in Table~\ref{tab:table3}, alongside data from the correlated configuration interaction calculations of Schmelcher et al \cite{Schmelcher1999, Schmelcher2000, Schmelcher2001} as well as Quantum Monte Carlo results due to \citet{Jones1999} and the recent results due to \citet{Schimeczek2013}. The absolute values of the binding energies are provided therein. In this study we investigated the three most tightly bound states within each parity sub-space. The corresponding weak field orbitals of these states are those listed in Table~\ref{tab:table1}. The values given in parentheses are from eigenvalue computations using spherical coordinates. These calculations were carried out using our pseudospectral atomic structure software developed in an earlier study \cite{HT2010}. An improved and faster version of the code was employed, again using the same levels of mesh refinement for maintaining consistency. The cylindrical pseudospectral method begins to lose accuracy as the magnetic field decreases, in the weak field region ($\beta_Z \stackrel{_<}{_\sim} 1 $), while in contrast, the spherical pseudospectral method loses accuracy in the upper end of the strong field regime, i.e., $\beta_Z \stackrel{_>}{_\sim} 1$. Therefore using a combination of the two types of codes, we can explore the entire range in $0 \leq \beta_Z \leq 1000$.


\begin{turnpage}
\begin{table}[h]
\centering
\caption{Absolute value of the binding energies of the positive parity states of helium. Energies are in units of Rydberg energies in the Coulomb potential of nuclear charge $Z=2$ for helium. Accurate data from other work is also provided for comparison. ($\beta_Z=\gamma/2Z^2$). The values given in parentheses are obtained from a faster version of our spherical atomic structure code developed earlier \cite{HT2010}.}
\begin{threeparttable}
\begin{tabular}{c@{\hspace{3mm}}c@{\hspace{3mm}}c@{\hspace{3mm}}c@{\hspace{3mm}}c@{\hspace{3mm}}c@{\hspace{3mm}}c@{\hspace{3mm}}c@{\hspace{3mm}}c@{\hspace{3mm}}c@{\hspace{3mm}}c@{\hspace{3mm}}c@{\hspace{3mm}}c@{\hspace{3mm}}}
\hline
\hline
& \multicolumn{4}{c}{$1^3(0)^+$} & \multicolumn{4}{c}{$1^3(-2)^+$} & \multicolumn{4}{c}{$1^3(-1)^+$} \\ 
\cline{2-5} \cline{6-9} \cline {10-13}\\
 & & Refs. & Refs. & Refs. & & Refs. & Refs. & Refs. & & Refs. & Refs. & Refs. \\
 $\beta_Z$ & Here & \cite{Ruder94}/\cite{TH2009} & \cite{Jones1999} & \cite{Schmelcher1999} & Here & \cite{Ruder94}/\cite{TH2009} & \cite{Jones1999} & \cite{Schmelcher2001}/\cite{Schimeczek2013} & Here & \cite{Ruder94}/\cite{TH2009} & \cite{Jones1999} & \cite{Schmelcher2000}/\cite{Ivanov2000}/\cite{Schimeczek2013} \\
\hline \\
0	& (1.0899) &        & 1.0871\tnote{d}	& 1.0876 & (1.0307) & 1.0179\tnote{a} &	        & 1.0278          & (1.0686) & 1.0668\tnote{a}	& 1.0657\tnote{c}  & 1.0666		\\
0.01	& (1.1256) & 1.1213 & 1.1213	        & 1.1220 & (1.0872) & 1.0852\tnote{a} & 1.0830	& 1.0833          & (1.1223) & 1.1183\tnote{a}	& 1.1177  	   & 1.1193		\\
 0.05	& (1.2103) & 1.1911 & 1.2056	        & 1.2064 & (1.2212) & 1.2175\tnote{a} & 1.2160	& 1.2167          & (1.2734) & 1.2691\tnote{a}	& 1.2683	   & 1.2704		\\
 0.1	& (1.2914) & 1.2133 & 1.2860	        & 1.2868 & (1.3495) & 1.3510\tnote{a} & 1.3436	& 1.3450          & (1.4209) & 1.4189\tnote{a}	& 1.4151	   & 1.4178		\\
 0.125	& (1.3301) & 	    & 		        & 1.3253 & (1.4063) & 	              & 	& 1.4016          & (1.4859) & 		        & 		   & 1.4828		\\
 0.2	& (1.4389) & 	    & 1.4330	        & 1.4338 & (1.5572) & 1.5598\tnote{a} & 	& 1.5525          & (1.6579) & 1.6585\tnote{a}	& 1.6508	   & 1.6544		\\
 0.25	& (1.5112) & 	    &		        & 1.4999 & (1.6459) & 	              &	        & 1.6411          & (1.7583) &  		& 		   & 1.7545		\\
 0.5	& 1.7722   & 1.4670 & 1.7718	        & 	 &  2.0035  & 2.0009\tnote{a} & 1.9945	&	          & (2.1582) & 2.1550\tnote{a}  & 2.1490	   & 		        \\
 0.625	& 1.8848   &	    &		        & 1.8841 &  2.1426  &	              &	        & 2.1425          & (2.3183) &		        &		   & 2.3128		\\
 0.7	& 1.9457   & 1.6690 & 1.9447	        & 	 &  2.2213  & 2.2246\tnote{a} & 2.2171	& 2.2210\tnote{c} & (2.4060) & 2.4029\tnote{a}  & 2.3956	   &		        \\
 1 	& 2.1607   & 1.9116 & 2.1601 	        & 	 &  2.4983  & 2.4981\tnote{a} & 2.4933	& 2.4978\tnote{c} & 2.7047   & 2.7026\tnote{a}	& 2.7000	   & 2.7043\tnote{c}	\\
 1.25   & 2.3141   & 	    &        	        & 2.3137 &  2.6943  & 	              & 	& 2.6940          & 2.9198   & 		        & 		   & 2.9197		\\
 2 	& 2.6863   & 2.4793 & 2.6840 	        & 	 &  3.1691  & 3.1685\tnote{a} & 3.1634	& 3.1691\tnote{c} & 3.4387   & 3.4384\tnote{a}	& 3.4333	   & 3.4386\tnote{c}	\\
 2.5 	& 2.8868   &        &        	        & 2.8862 &  3.4275  &                 & 	& 3.4274          & 3.7203   & 		        & 		   & 3.7203		\\
 5 	& 3.6311   & 3.4672 & 3.6272 	        & 	 &  4.3785  & 4.3740\tnote{a} & 4.3693	& 4.3783\tnote{c} & 4.7539   & 4.7502\tnote{a}	& 4.7441           & 4.7523\tnote{c}	\\
 6.25 	& 3.9082   &        &         	        & 3.9076 &  4.7383  & 	              & 	& 4.7380          & 5.1425   & 		        & 		   & 5.1421		\\
 7 	& 4.0618   &        &        	        & 	 &  4.9310  & 4.9268\tnote{a} & 4.9203	& 4.9310\tnote{c} & 5.3514   & 5.3474\tnote{a}	& 5.3408	   & 5.3507\tnote{c}	\\
10 	& 4.5742   & 4.4362 & 4.5693 	        & 	 &  5.5856  & 5.5851\tnote{a} & 5.5770	& 5.5899\tnote{c} & 6.0548   & 6.0543\tnote{a}	& 6.0506	   & 6.0624\tnote{c}	\\
12.5 	& 4.9219   &        &        	        & 4.9215 &  6.0441  & 	              & 	& 6.0443          & 6.5527   & 	 	        & 		   & 6.5524		\\
20 	& 5.7538   & 5.6367 & 5.7473 	        & 	 &  7.0968  & 6.9867          & 7.0938	& 7.1120\tnote{c} & 7.6863   & 7.5750		& 7.6845	   & 7.7018\tnote{c}	\\
25 	& 6.1896   & 	    & 	 	        & 	 &  7.6563  & 	              & 	&                 & 8.2896   & 		        & 		   & 8.2895\tnote{b}	\\
50 	& 7.7403   & 7.6473 & 7.7334 	        & 	 &  9.6683  & 9.5806          & 9.6643	& 9.6949\tnote{c} & 10.4447  & 10.3567	        & 10.4438	   & 10.4726\tnote{c}	\\
62.5 	& 8.3074   & 	    & 	 	        & 	 &  10.4022 & 	              & 	& 	          & 11.2348  & 		        & 		   & 11.2333\tnote{b}	\\
70 	& 8.6085   & 	    & 	 	        & 	 &  10.8018 & 	              & 10.7977	& 10.8306\tnote{c}& 11.6577  & 		        & 11.6533	   & 11.6879\tnote{c}	\\
100 	& 9.6191   & 9.5434 & 9.6143 	        & 	 &  12.1214 & 12.0449         & 12.1142	& 12.1584\tnote{c}& 13.0635  & 12.9904	        & 13.0632	   & 13.1055\tnote{c}   \\
125 	& 10.3040  & 	    &        	        & 	 &  13.0077 & 	              & 	& 	          & 14.0163  & 		        & 		   & 14.0161\tnote{b}	\\
200 	& 11.8785  & 11.8111&        	        & 	 &  15.0689 & 15.0117         & 	& 	          & 16.2018  & 16.1516	        & 		   & 		        \\
250 	& 12.6878  & 	    &        	        & 	 &  16.1375 & 	              & 	& 	          & 17.3497  & 		        & 		   & 17.3495\tnote{b}	\\
500 	& 15.5033  & 15.4537&        	        & 	 &  19.8595 & 19.5816         & 	& 	          & 21.2918  & 21.2521	        & 		   & 		        \\
700 	& 17.0409  & 	    &        	        & 	 &  21.8937 & 	              & 	& 	          & 23.4470  & 		        & 		   & 		        \\
1000	& 18.7920  & 18.7519&        	        & 	 &  24.2400 & 24.2004         & 	& 	          & 25.9295  & 25.8917	        & 		   &  		        \\
\hline
\hline
\end{tabular}
\begin{tablenotes}
       \item[a] Ref.~\cite{TH2009}
       \item[b] Ref.~\cite{Ivanov2000}
       \item[c] Ref.~\cite{Schimeczek2013}
       \item[d] Ref.~\cite{Jones1996}
\end{tablenotes}
\end{threeparttable}
\label{tab:table2}
\end{table}
\end{turnpage}
It can be seen upon examining the data in Tables~\ref{tab:table2} and \ref{tab:table3}, that the fully converged results obtained in the current study are in good agreement with values obtained elsewhere, given that the current study is a single configuration calculation. However the recent results due to \citet{Schimeczek2013} are on average about $\Delta \approx 0.22\%$ more bound for $\beta_Z \stackrel{_>}{_\sim} 10$. The positive parity states of helium are more bound than the negative parity states in intense magnetic field strengths. For the positive parity states in Table~\ref{tab:table2}, over the entire range of magnetic field strengths investigated, our estimates of the binding energies agree with estimates elsewhere \cite{Jones1999, Schmelcher1999, Schmelcher2000, Schmelcher2001, Schimeczek2013} to on average $\Delta \approx 0.19\%$, $0.23\%$ and $0.12\%$, for the states $1^3(0)^+$, $1^3(-2)^+$ and $1^3(-1)^+$, respectively. We noticed loss of accuracy of the cylindrical pseudospectral method in the lower magnetic field regime ($\beta_Z \stackrel{_<}{_\sim}$ 1) and therefore employed our spherical code for the lower magnetic field strengths while in the range $1 \leqslant \beta_Z \leqslant 1000$ we employed the cylindrical code. We observed that the cylindrical code (and the extrapolation method described above) maintained accuracy to within $10^{-6}$ to $10^{-5}E_{Z,\infty}$ in this latter range. In Table~\ref{tab:table3} the agreement of our results with those of Refs.~\cite{Jones1999, Schmelcher1999, Schmelcher2000, Schmelcher2001, Schimeczek2013} is $\Delta \approx 0.15\%$, $0.21\%$ and $0.14\%$, for the negative parity states $1^3(0)^-$, $1^3(-1)^-$ and $1^3(-2)^-$, respectively.

One of the aims of the current study is to provide a fast method for the calculation of the energy landscape of atoms in intense magnetic fields; therefore, we have additionally calculated fits to the data provided in Tables~\ref{tab:table2} and \ref{tab:table3}. The model fits are rational functions whose analytic form is given by,
\begin{equation}
f(x) = \frac{\sum_{i=0}^n a_i x^i}{x^{m} + \sum_{i=0}^{m-1} b_i x^i},
\label{eq:68}
\end{equation}
where $x=\ln(1+\beta_Z)$ and $m=n-2$. The fitting was carried out using a non-linear least squares Levenberg-Marquardt algorithm with line searches \cite{NR1992}. The coefficients and the maximal fitting errors over the entire range $\beta_Z=0$ to $\beta_Z=10^3$ are given in Table~\ref{tab:table4}. These fitting functions could be employed directly in atmosphere models of neutron stars rather than incorporating a code that calculates the binding energy. Thus, atmosphere models which are computationally intensive to begin with, need not be further complicated with the addition of an atomic structure calculation module, even though the software developed in this study is compact and computationally efficient.

\subsection{\label{Li_results} The Lithium Atom}

We investigated the six most tightly bound states of the lithium atom in intense magnetic fields. The binding energies obtained by solving the eigenvalue problem are shown in Tables~\ref{tab:table5} and \ref{tab:table6}. These tables show the results for the positive and negative parity states, respectively. As in the case of the helium atom, the HF binding energies are results that were obtained after extrapolating to the limit of infinite domain size. 

In contrast to the helium atom, lithium has been investigated far less in the literature, and data is scarce for the binding energies of the different states, particularly in the intense field regime for the negative parity states.
Table~\ref{tab:table5} shows the converged HF binding energies for the three most tightly bound states in the positive parity sub-space of the lithium atom. Over the entire range of magnetic field strengths investigated in the current single configuration study, the results obtained here agree with results obtained elsewhere \cite{Schmelcher_lithium2004, Ivanov2000, Ivanov1998, Schimeczek2013} to on average $\Delta \approx 0.29\%$, $0.5\%$ and $0.58\%$ for the positive parity states $1^4(-2)^+$, $1^4(-1)^+$ and $1^4(-3)^+$, respectively. 

Table~\ref{tab:table6} shows the converged HF binding energies of the negative parity states of lithium. Of the three most tightly bound states in this parity sub-space, only one has been investigated in the literature so far; the state $1^4(-1)^-$. The average agreement between our results and those of \citet{Schmelcher_lithium2004, Ivanov2000} is on average $\Delta \approx 0.51\%$, over the entire range of magnetic field strengths considered in this study. We observed that the extrapolation to the limit of infinite domain size maintained accuracy to within $10^{-5}$ to $10^{-6} E_{Z,\infty}$ over the entire range of magnetic field strengths. Moreover, since our results are upper bounds for the uncorrelated single-configuration binding energies, the data for the states $1^4(-2)^-$ and $1^4(-3)^-$ given in Table~\ref{tab:table6} can be improved using post-HF methods. The $2-$D wave functions of these states calculated herein could be employed as initial estimates for improvements. 

Moreover, with increasing magnetic field strength in the intense magnetic field regime, the most tightly bound negative parity state of lithium is seen to be the state $1^4(-2)^-$, which is comprised of the field-free orbitals $1s_02p_{-1}3d_{-1}$. This crossover occurs at around $\beta_Z \approx 10$, below this field the $1^4(-1)^-$ is the most tightly bound of the three negative parity states of lithium shown in Table~\ref{tab:table6}. To the best of our knowledge, this crossover has not been reported elsewhere in the literature. In addition, it can be seen by comparing the binding energies reported in Tables~\ref{tab:table5} and \ref{tab:table6}, that the state $1^4(-2)^-$, is also among the most tightly bound states of lithium in intense magnetic fields. Moreover, the third state shown therein, the $1^4(-2)^-$ state, comprised of the orbitals $1s_02p_{-1}3d_{-1}$, also has not been investigated in the literature. This latter state also becomes tightly bound with increasing magnetic field strength in the intense field regime. 

Thus overall we see that for the six most tightly bound states of lithium in intense magnetic fields, two of the states have not been investigated earlier at all ($1^4(-3)^-$ and $1^4(-2)^-$) and a third state ($1^4(-2)^+$) has not been investigated in the intense field regime ($\beta_Z \geqslant 0.5556$). Therefore the results presented here appear to be the first of such studies and represent upper bounds to the uncorrelated single-configuration binding energies. In addition we see that for the remaining three states that were investigated, the binding energies obtained in the current study are in relative good agreement with estimates obtained elsewhere at the $ \stackrel{_<}{_\sim} 0.5 \%$ level overall.

Furthermore, for the sake of facilitating atmosphere and crustal models of neutron stars, we have also carried out rational function fits to the data. Once again these analytic forms can be implemented directly in such codes, thereby circumventing the need for atomic structure calculations altogether. The rational functions have the same functional form as those described in Eq.~(\ref{eq:68}) above. The coefficients of these rational functions are given in Table~\ref{tab:table7} alongside estimates of the fitting errors. 

In the following section, we summarize the findings alongside a brief discussion of further avenues for investigation.
\begin{turnpage}
\begin{table}[h]
\centering
\caption{Absolute value of the binding energies of the negative parity states of helium. Energies are in units of Rydberg energies in the Coulomb potential of nuclear charge $Z=2$ for helium. Accurate data from other work is also provided for comparison. ($\beta_Z=\gamma/2Z^2$).}
\begin{threeparttable}
\begin{tabular}{c@{\hspace{3mm}}c@{\hspace{3mm}}c@{\hspace{3mm}}c@{\hspace{3mm}}c@{\hspace{3mm}}c@{\hspace{3mm}}c@{\hspace{3mm}}c@{\hspace{3mm}}c@{\hspace{3mm}}c@{\hspace{3mm}}c@{\hspace{3mm}}c@{\hspace{3mm}}c@{\hspace{3mm}}}
\hline
\hline
& \multicolumn{4}{c}{$1^3(0)^-$} & \multicolumn{4}{c}{$1^3(-1)^-$} & \multicolumn{4}{c}{$1^3(-2)^-$} \\ 
\cline{2-5} \cline{6-9} \cline {10-13}\\
 & & Refs. & Refs. & Refs. & & Refs. & Refs. & Refs. & & Refs. & Refs. & Refs. \\
 $\beta_Z$ & Here & \cite{Ruder94}/\cite{TH2009} & \cite{Jones1999} & \cite{Schmelcher1999}/\cite{Schimeczek2013} & Here & \cite{Ruder94}/\cite{TH2009} & \cite{Jones1999} & \cite{Schmelcher2000} & Here & \cite{Ruder94}/\cite{TH2009} & \cite{Jones1999} & \cite{Schmelcher2001}\\
\hline \\
 0	& (1.0686) & 1.0641\tnote{a} & 1.0657\tnote{c}  & 1.0665	          & (1.0307) &        &          & 1.0278 & (1.0189)  & 	& 		& 1.0156	\\
 0.01	& (1.1056) & 1.1029\tnote{a} & 1.1016	   	& 1.1026	          & (1.0745) & 1.0693 & 1.0702   & 1.0705 & (1.0643)  & 1.0579	& 1.0601	& 1.0603	\\
 0.05	& (1.2147) & 1.2135\tnote{a} & 1.2099		& 1.2112	          & (1.1808) & 1.1565 & 1.1756   & 1.1761 & (1.1684)  & 1.1364	& 1.1634	& 1.1636	\\
 0.1	& (1.3229) & 1.3231\tnote{a} & 1.3174		& 1.3191	          & (1.2846) & 1.2270 & 1.2764   & 1.2795 & (1.2709)  & 1.1979	& 1.2650	& 1.2655	\\
 0.125	& (1.3727) & 		     &			& 1.3669	          & (1.3308) &        & 	 & 1.3255 & (1.3166)  & 	& 		& 1.3110	\\
 0.2	& (1.4982) & 1.4988\tnote{a} & 1.4914		& 1.4936~[1.4938]\tnote{c} & (1.4539) & 1.3174 & 1.4469   & 1.4481 & (1.4384)  & 1.2723	& 1.4319	& 1.4326	\\
 0.25	& (1.5721) & 		     &			& 1.5676~[1.5677]\tnote{c} & (1.5262) &        &	         & 1.5202 & (1.5044)  & 	& 		& 1.5042	\\
 0.5	&  1.8648  & 1.8647\tnote{a} & 1.8593		& 		          &  1.8131  & 1.5131 & 1.8078   &	  &  1.7918   & 1.4937	& 1.7912	& 		\\
 0.625	&  1.9798  &		     & 			& 1.9796	          &  1.9263  &	      &	         & 1.9259 &  1.9092   &		& 		& 1.9090	\\
 0.7	&  2.0465  & 2.0461\tnote{a} & 2.0404		& 		          &  1.9935  & 1.7202 & 1.9877   &	  &  1.9717   & 1.7012	& 1.9709	&		\\
 1 	&  2.2689  & 2.2685\tnote{a} & 2.2641		& 		          &  2.2159  & 1.9678 & 2.2097   & 	  &  2.1939   & 1.9493	& 2.1931	& 		\\
 1.25 	&  2.4245  & 		     & 			& 2.4243	          &  2.3689  &        & 	 & 2.3687 &  2.3524   & 	& 		& 2.3522        \\
 2 	&  2.8069  & 2.8060\tnote{a} & 2.7989		& 		          &  2.7502  & 2.5433 & 2.7441   & 	  &  2.7295   & 2.5269	& 2.7287	&  		\\
 2.5 	&  3.0059  & 		     & 			& 3.0057~[3.0058]\tnote{c} &  2.9512  &        & 	 & 2.9511 &  2.9364   & 	& 		&  2.9361       \\
 5 	&  3.7524  & 3.7522\tnote{a} & 3.7466		& 		          &  3.7004  & 3.5369 & 3.6950   & 	  &  3.6836   & 3.5242	& 3.6827	&  		\\
 6.25 	&  4.0297  & 		     & 			& 4.0297	          &  3.9799  &        & 	 & 3.9795 &  3.9684   & 	& 		&  3.9680	\\
 7 	&  4.1819  & 4.1817\tnote{a} & 4.1762		& 		          &  4.1315  &        & 4.1264   & 	  &  4.1163   & 	& 4.1155	&  		\\
10 	&  4.6921  & 4.6920\tnote{a} & 4.6862		& 		          &  4.6434  & 4.5067 & 4.6387   &	  &  4.6304   & 4.4971	& 4.6275	& 		\\
12.5 	&  5.0406  & 		     & 			& 5.0400~ [5.0401]\tnote{c} &  4.9947  &        & 	 & 4.9947 &  4.9866   & 	& 		& 4.9860	\\
20 	&  5.8589  & 5.7495	     & 5.8585		& 		          &  5.8195  & 5.7056 & 5.8158   &	  &  5.8103   & 5.6989	& 5.8091	&  		\\
25 	&  6.2942  & 		     & 			& 6.2940\tnote{b}         &  6.2545  &        &	         &	  &  6.2462   & 	& 		&   		\\
50 	&  7.8349  & 7.7491	     & 7.8346		& 		          &  7.8016  & 7.7118 & 7.7958   &	  &  7.7962   & 7.7080	& 7.7940	&  		\\
62.5 	&  8.3991  & 		     & 			& 		          &  8.3682  &        & 	 &	  &  8.3632   & 	& 		&   		\\
70 	&  8.7004  & 		     & 8.6998		& 		          &  8.6684  &        & 8.6651   &	  &  8.6646   & 	& 8.6620	&  		\\
100 	&  9.7078  & 9.6370	     & 9.7075		& 		          &  9.6799  & 9.6039 & 9.6748   &	  &  9.6756   & 9.6015	& 9.6724	&  		\\
125 	&  10.3885 & 		     & 			& 		          &  10.3617 &        & 	 &	  &  10.3586  & 	& 		&  		\\
200 	&  11.9501 & 11.8970	     & 			& 		          &  11.9339 & 11.8675& 	 &	  &  11.9330  & 11.8661	& 		&  		\\
250 	&  12.7593 & 		     & 			& 		          &  12.7427 &        & 	 &	  &  12.7398  & 	& 		&  		\\
500 	&  15.5573 & 15.5307	     & 			& 		          &  15.5449 & 15.5050& 	 &	  &  15.5290  & 15.5043	& 		&  		\\
700 	&  17.0580 & 		     & 			& 		          &  17.0524 &        & 	 &	  &  17.0427  & 	& 		&  		\\
1000 	&  18.8466 & 18.8230	     & 			& 		          &  18.8275 & 18.7997& 	 &	  &  18.8198  & 18.7993	& 		&  		\\
\hline
\hline
\end{tabular}
\begin{tablenotes}
       \item[a] Ref.~\cite{TH2009}
       \item[b] Ref.~\cite{Ivanov2000}
       \item[c] Ref.~\cite{Schimeczek2013}
\end{tablenotes}
\end{threeparttable}
\label{tab:table3}
\end{table}
\end{turnpage}

\begin{table}[ht]
  \caption{Coefficients of the different rational
    functions for fitting the six states of helium discussed.  The absolute
    maximum fractional error of the eigenvalue relative to the fit from
    $\beta_Z=0$ to $\beta_Z=10^3$ is reported in the variable $\epsilon$}

\begin{tabular}{clcl}
\hline
\hline
State & \multicolumn{1}{c}{Coefficients} &
State & \multicolumn{1}{c}{Coefficients} \\
\hline
$\begin{array}{cr}
1^3(-1)^+ \\
1{\rm s}_02{\rm p}_{-1} 
\end{array}$ & 
$\begin{array}{cr}
  a_0=&       -7.0472899\\
  a_1=&       -12.1364585\\
  a_2=&       14.9186598\\
  a_3=&       -1.8780419\\
  a_4=&       0.7447550\\  
  b_0=&       -6.1146090\\
  b_1=&       3.4787694\\
  \epsilon=& 3 \times10^{-2}
\end{array}$ &
$\begin{array}{c}
1^3(-2)^- \\
1{\rm s}_04{\rm f}_{-2} 
\end{array}$ & 
$\begin{array}{cr}
  a_0=&       5.1971696\\
  a_1=&       10.0283242\\
  a_2=&       -0.5210301\\
  a_3=&       0.5165584\\
  b_0=&       4.8072912\\
  \epsilon=& 1 \times10^{-2}
\end{array}$ \\
\hline
$\begin{array}{c}
1^3(-2)^+ \\
1{\rm s}_03{\rm d}_{-2} 
\end{array}$ & 
$\begin{array}{cr}
  a_0=&      19.7181974\\
  a_1=&      156.4098592\\
  a_2=&      115.7638407\\
  a_3=&      1.8563608\\
  a_4=&      3.3370287\\  
  b_0=&      18.9302257\\
  b_1=&      78.9859850\\
  \epsilon=& 5 \times10^{-4}
\end{array}$ & 
$\begin{array}{c}
1^3(-1)^- \\
1{\rm s}_03{\rm d}_{-1} 
\end{array}$ & 
$\begin{array}{cr}
  a_0=&       -2.1093928\\
  a_1=&       1.2121356\\
  a_2=&       10.3242855\\
  a_3=&       -0.7599149\\
  a_4=&       0.5200099\\    
  b_0=&       -1.9301991\\
  b_1=&       4.4132323\\   
  \epsilon=& 1 \times10^{-2}
\end{array}$ \\
\hline
$\begin{array}{c}
1^3(0)^+ \\
1{\rm s}_02{\rm s}_{0} 
\end{array}$ & 
$\begin{array}{cr}
  a_0=&    25.4217352\\
  a_1=&    115.9813417\\
  a_2=&    64.7201003\\
  a_3=&    2.5281290\\
  a_4=&    1.8168572\\
  b_0 =&   23.1414290\\
  b_1 =&   58.2065631\\
  \epsilon=& 2 \times 10^{-4}
\end{array}$ & 
$\begin{array}{c}
1^3(0)^- \\
1{\rm s}_02{\rm p}_{0} 
\end{array}$ & 
$\begin{array}{cr}
  a_0=&    -3.1828923\\
  a_1=&    -0.3432482\\
  a_2=&    10.3960349\\
  a_3=&    -0.8278049\\
  a_4=&    0.5141009\\
  b_0 =&   -2.7702334\\
  b_1=&	   4.2104702\\
  \epsilon=& 2 \times 10^{-2}
\end{array}$ \\
\hline
\hline
\end{tabular}
\label{tab:table4}
\end{table}

\begin{table}[h]
\centering
\caption{Absolute value of the binding energies of the positive parity states of lithium. Energies are in units of Rydberg energies in the Coulomb potential of nuclear charge $Z=3$ for lithium. Accurate data from other work is also provided for comparison. ($\beta_Z=\gamma/2Z^2$).}
\begin{threeparttable}
\begin{tabular}{c@{\hspace{3mm}}c@{\hspace{3mm}}c@{\hspace{3mm}}c@{\hspace{3mm}}c@{\hspace{3mm}}c@{\hspace{3mm}}c@{\hspace{3mm}}}
\hline
\hline
& \multicolumn{2}{c}{$1^4(-2)^+$} & \multicolumn{2}{c}{$1^4(-1)^+$} & \multicolumn{2}{c}{$1^4(-3)^+$} \\ 
\cline{2-3} \cline{4-5} \cline {6-7}\\
 $\beta_Z$ & Here & Ref.~\cite{Schmelcher_lithium2004} & Here & Elsewhere & Here & Elsewhere  \\
\hline
0	&   (1.1492) & 1.1491	& (1.1968)	& 1.1926\tnote{a}	& (1.1357) &  1.1427\tnote{a}~ [1.1299]\tnote{d} 	\\
0.00056	&   (1.1541) & 1.1544	& (1.2024)	& 1.1969\tnote{a}	& (1.1425) &  1.1487\tnote{a}				\\
0.0028	&   (1.1780) & 1.1720	& (1.2194)	& 1.2121\tnote{c}	& (1.1652) &  1.1663\tnote{a}				\\
0.0056	&   (1.1961) & 1.1901	& (1.2390)	& 1.2334\tnote{a}	& (1.1897) &  1.1869\tnote{a}				\\		
0.0111	&   (1.2267) & 1.2203	& (1.2735)	& 1.2674\tnote{a}	& (1.2324) &  1.2278\tnote{a}				\\
0.0278	&   (1.2964) & 1.2886	& (1.3530)	& 1.3463\tnote{a}	& (1.3354) &  1.3294\tnote{a}				\\
0.0556	&   (1.3912) & 1.3930	& (1.4529)	& 1.4432\tnote{a}	& (1.4699) &  1.4627\tnote{a}				\\
0.5 	&    2.1745  & 		&  2.3607	& 			& 2.5903   &   						\\
0.5556 	&    2.4146  & 2.4145	&  2.4284	& 2.4280\tnote{a} 	& 2.6579   &  2.6572\tnote{a}				\\
1 	&    2.7158  &		&  2.9551	& 			& 3.2818   &  						\\
1.1111 	&    2.8124  &		&  3.0619	& 3.0432\tnote{b}	& 3.4056   &  3.3695\tnote{b}				\\
2 	&    3.4404  &		&  3.7473	& 			& 4.1986   &  						\\
2.3636  &            &          &               &                       & 4.4581   &  4.4133\tnote{b}~ [4.4203]\tnote{e}        \\
2.7778 	&    3.8590  &		&  4.2030	& 4.1781\tnote{b}	& 4.7247   &  4.6779\tnote{b}   			\\
5 	&    4.7467  &		&  5.1669	& 			& 5.8362   &  						\\
5.5556 	&    4.9262  &		&  5.3616	& 5.3304\tnote{c}	& 6.0606   &  6.0043\tnote{b}				\\
7 	&    5.3444  &		&  5.8140	& 			& 6.5822   &  						\\
10 	&    6.0571  &		&  6.5846	& 			& 7.4703   &  						\\
11.1111 &    6.2835  &		&  6.8301	& 6.7909\tnote{c}	& 7.7532   &  7.6856\tnote{c}				\\
11.8178 &            &          &               &                       & 7.9232   &  7.8544\tnote{b}~ [7.8711]\tnote{e}        \\
20 	&    7.7046  &		&  8.3606	& 			& 9.5172   &  						\\
23.6356 &            &          &               &                       & 10.0807  &  9.9989\tnote{b}~ [10.0238]\tnote{e}       \\
27.7778 &   8.6194   &		&  9.3450       & 9.2936\tnote{c}	& 10.6480  &  10.5685\tnote{b}				\\
50 	&   10.5017  &		&  11.3643      & 			& 12.9842  &  						\\
55.5556 &   10.8742  &		&  11.7638      & 11.7000\tnote{c}	& 13.4456  &  13.3464\tnote{b}   			\\
70 	&   11.7297  &		&  12.6825      & 			& 14.5023  &  						\\
100 	&   13.1685  &		&  14.2197      & 			& 16.2849  &  						\\
118.1780&            &          &               &                       & 17.1804  &  17.0622\tnote{b}~ [17.1231]\tnote{e}      \\
200 	&   16.3929  &		&  17.6559      & 			& 20.2679  & 						\\
277.7778&   18.1335  &		&  19.5079      & 			& 22.4181  &  22.2774\tnote{b}				\\
500 	&   21.6227  &		&  23.2111      & 			& 26.7255  & 						\\
555.5556&   22.2335  &		&  23.9240      & 			& 27.5636  &  27.4029\tnote{b} 				\\
700 	&   23.8351  &		&  25.5639      & 			& 29.4767  & 						\\
1000 	&   26.3906  &		&  28.2108      & 			& 32.6287  &  						\\
\hline 
\hline
\end{tabular}
\begin{tablenotes}
       \item[a] Ref.~\cite{Schmelcher_lithium2004}
       \item[b] Ref.~\cite{Ivanov2000}
       \item[c] Ref.~\cite{Ivanov1998}
       \item[d] Ref.~\cite{Ralchenko2008}, experimental result.
       \item[e] Ref.~\cite{Schimeczek2013}
\end{tablenotes}
\end{threeparttable}
\label{tab:table5}
\end{table}
\begin{table}[h]
\centering
\caption{Absolute value of the binding energies of the negative parity states of lithium. Energies are in units of Rydberg energies in the Coulomb potential of nuclear charge $Z=3$ for lithium. Accurate data from other work is also provided for comparison. ($\beta_Z=\gamma/2Z^2$).}
\begin{threeparttable}
\begin{tabular}{c@{\hspace{3mm}}c@{\hspace{3mm}}c@{\hspace{3mm}}c@{\hspace{3mm}}c@{\hspace{3mm}}}
\hline
\hline
& \multicolumn{2}{c}{$1^4(-1)^-$} & \multicolumn{1}{c}{$1^4(-2)^-$} & \multicolumn{1}{c}{$1^4(-3)^-$} \\ 
\cline{2-2} \cline{3-4} \cline {5-5}\\
 $\beta_Z$ & Here & Elsewhere & Here & Here  \\
\hline
0	 &  (1.1687)  & 1.1652\tnote{c}	& (1.1400) 	& (1.1306)	\\
0.00056	 &  (1.1735)  & 1.1695\tnote{c}	& (1.1457)	& (1.1361)	\\
0.0028	 &  (1.1909)  & 1.1865\tnote{c}	& (1.1644)	& (1.1556)	\\
0.0056	 &  (1.2110)  & 1.2065\tnote{c}	& (1.1827)	& (1.1755)	\\		
0.0111	 &  (1.2469)  & 1.2417\tnote{c}	& (1.2183)	& (1.2104)	\\
0.0278	 &  (1.3354)  & 1.3297\tnote{c}	& (1.3073)	& (1.2964)	\\
0.0556	 &  (1.4500)  & 1.4463\tnote{c}	& (1.4248)	& (1.4113)	\\
 0.5 	 &  2.4358    & 	    	& 2.4048   	& 2.3804 	\\
 0.5556  &  2.4899    & 2.4898\tnote{c} & 2.4853   	& 2.4605 	\\
 1 	 &  3.0313    &     		& 3.0074   	& 2.9807 	\\
 1.1111  &  3.1376    & 3.1035\tnote{a} & 3.1150   	& 3.0881 	\\
 2 	 &  3.8180    & 		& 3.8029   	& 3.7757 	\\
 2.7778  &  4.2693    & 4.2319\tnote{a}	& 4.2585   	& 4.2316 	\\
 5 	 &  5.2238    & 		& 5.2196  	& 5.1945 	\\
 5.5556  &  5.4168    & 5.3767\tnote{b}	& 5.4136   	& 5.3889 	\\
 7 	 &  5.8654    &     		& 5.8641   	& 5.8405 	\\
10 	 &  6.6301    &    		& 6.6313   	& 6.6095 	\\
11.1111  &  6.8738    & 6.8298\tnote{b}	& 6.8757   	& 6.8544 	\\
20 	 &  8.3958    &    		& 8.3998   	& 8.3819 	\\
27.7778  &  9.3762    & 9.3242\tnote{a}	& 9.3808   	& 9.3646 	\\
50 	 &  11.3892   &  		& 11.3941 	& 11.3809	\\
55.5556  &  11.7877   & 11.7269\tnote{a}& 11.7922 	& 11.7799	\\
70 	 &  12.7043   &  		& 12.7090 	& 12.6974	\\
100 	 &  14.2387   & 		& 14.2445 	& 14.2331	\\
200 	 &  17.6700   &  		& 17.6795 	& 17.6601	\\
277.7778 &  19.5196   &   		& 19.5324 	& 19.5173	\\
500 	 &  23.2201   &    		& 23.2339 	& 23.2195	\\
555.5556 &  23.9385   &     		& 23.9592 	& 23.9414	\\
700 	 &  25.5784   &      		& 25.6165 	& 25.5923	\\
1000 	 &  28.2764   &      		& 28.3062 	& 28.2948       \\
\hline 
\hline
\end{tabular}
\begin{tablenotes}
       \item[a] Ref.~\cite{Ivanov1998}
       \item[b] Ref.~\cite{Ivanov2000}
       \item[c] Ref.~\cite{Schmelcher_lithium2004}
\end{tablenotes}
\end{threeparttable}
\label{tab:table6}
\end{table}

\begin{table}[t]
  \caption{Coefficients of the different rational
    functions for fitting the six states of lithium discussed.  The absolute
    maximum fractional error of the eigenvalue relative to the fit from
    $\beta_Z=0$ to $\beta_Z=10^3$ is reported in the variable $\epsilon$}

\begin{tabular}{clcl}
\hline
\hline
State & \multicolumn{1}{c}{Coefficients} &
State & \multicolumn{1}{c}{Coefficients} \\
\hline
$\begin{array}{cr}
1^4(-3)^+ \\
1{\rm s}_02{\rm p}_{-1}3{\rm d}_{-2} 
\end{array}$ & 
$\begin{array}{cr}
  a_0=&    5.1377099\\
  a_1=&    16.4802171\\ 
  a_2=&	   -1.3253446\\
  a_3=&    0.9358112\\ 
  b_0=&     4.2869346\\
  \epsilon=& 7 \times 10^{-2}
\end{array}$ &
$\begin{array}{c}
1^4(-3)^- \\
1{\rm s}_02{\rm p}_{-1}4{\rm f}_{-2} 
\end{array}$ & 
$\begin{array}{cr}
  a_0=&    14.3494286\\
  a_1=&    153.2018162\\ 
  a_2=&    118.5582675\\
  a_3=&    1.5032372\\ 
  a_4=&    3.3925135\\ 
  b_0=&    12.6003601\\
  b_1=&    67.7819714\\
   \epsilon=& 6 \times 10^{-4}
\end{array}$ \\
\hline
$\begin{array}{c}
1^4(-1)^+ \\
1{\rm s}_02{\rm s}_02{\rm p}_{-1} 
\end{array}$ & 
$\begin{array}{cr}
  a_0=&    -0.2858217\\
  a_1=&    5.1103556\\ 
  a_2=&	   14.7469530\\
  a_3=&    -0.9845303\\ 
  a_4=&    0.8027005\\
  b_0=&    -0.2312447\\
  b_1=&    4.6481310\\
   \epsilon=& 2 \times 10^{-2}
\end{array}$ & 
$\begin{array}{c}
1^4(-1)^- \\
1{\rm s}_02{\rm p}_{0}2{\rm p}_{-1} 
\end{array}$ & 
$\begin{array}{cr}
  a_0=&    9.8120757\\
  a_1=&    101.3402326\\ 
  a_2=&	   76.6650758\\
  a_3=&    0.9558892\\ 
  a_4=&    2.3667694\\ 
  b_0=&    8.3449537\\
  b_1=&    43.4708070\\
   \epsilon=& 1 \times 10^{-3}
\end{array}$ \\
\hline
$\begin{array}{c}
1^4(-2)^+ \\
1{\rm s}_02{\rm s}_{0}3{\rm d}_{-2} 
\end{array}$ & 
$\begin{array}{cr}
  a_0=&    -7.8516707\\
  a_1=&    -12.5891289\\
  a_2=&    15.2720426\\
  a_3=&    -2.0240435\\
  a_4=&    0.7777919\\
  b_0 =&   -6.5356712\\
  b_1=&	   3.6017700\\
  \epsilon=& 7 \times 10^{-2}
\end{array}$ & 
$\begin{array}{c}
1^4(-2)^- \\
1{\rm s}_02{\rm p}_{-1}3{\rm d}_{-1} 
\end{array}$ & 
$\begin{array}{cr}
  a_0=&    -2.5236564\\
  a_1=&    -1.8911941\\
  a_2=&    15.4669138\\
  a_3=&    -1.5653288\\
  a_4=&    0.8148882\\
  b_0=&    -2.1270061\\
  b_1=&    3.9703734\\
   \epsilon=& 4 \times 10^{-2}
\end{array}$ \\
\hline
\hline
\end{tabular}
\label{tab:table7}
\end{table}

\section{\label{sec:conclusion} Conclusion}

%

In the current study we have investigated low-Z atoms, helium and lithium in intense magnetic fields. A two-dimensional single-configuration Hartree-Fock method, as described in Ref.~\cite{TH2009}, was adopted. A key feature of the method is that the potentials for the inter-electronic interactions are obtained as solutions to the elliptic partial differential equations as given in Eqs.~(\ref{eq:3}) and (\ref{eq:4}). The HF equations in Eq.~(\ref{eq:1}) are solved using the self consistent field method. The system size grew as $n_e(N-1)^2 \times n_e(N-1)^2$, with $n_e$ the number of electrons in the coupled problem and $N$ the number of grid points in each direction. 

A pseudospectral approach was adopted for the numerical solution of the problem using cylindrical coordinates so as to facilitate calculations in the intense field regime. Domain discretization was achieved using the commonly employed Chebyshev-Lobatto spectral collocation method. The resulting discretized and coupled eigenvalue problem problem was solved using standard sparse matrix methods using the software package ARPACK. The key enabling advantage of the psuedospectral approach is the immensely reduced computational time particularly, since we have adopted here an unrestricted two-dimensional approach to the problem \cite{TH2009}. The latter has the advantage that it does not require a basis of functions to describe the wave functions. Thus the wave functions obtained in the current unrestricted 2D approach can effectively be thought of as those arising from the superposition of a large number of basis functions. 

We presented data for the six most tightly bound states of the helium atom in intense magnetic fields, in Tables~\ref{tab:table2} and \ref{tab:table3}. These were seen to be consistent with findings elsewhere. Similarly we investigated the six most tightly bound states of the lithium atom as well. However, we found that the data in the literature to be rather scarce for lithium. As a result we could only compare our results for four of the six states characterized in this study. We obtained, apparently for the first time, calculations for the binding energies for the states $1^4(-2)^-$ and $1^4(-3)^-$ of lithium. We find that the the $1^4(-2)^-$ state is also the most tightly bound negative parity state of lithium in the limit of intense magnetic fields. 

The work described herein was motivated primarily by the need to have accurately determined upper bounds for the binding energies of atoms in intense magnetic fields employing a computationally straight-forward implementation. As the atomic structure software developed here is compact and computationally economical, it produces accurate results within a short amount of computing time. As a result, it can be incorporated directly into atmosphere and crustal models for neutron stars. However, while this may be desirable, it may present an additional layer of computational complexity. The user may wish to circumvent this by employing the rational function fits given in Tables~\ref{tab:table4} and \ref{tab:table7}. These analytic forms, model the data in the range $0 \leq \beta_Z \leq1000$ and thus may simplify atmosphere and crustal models considerably. Estimates of binding energies and oscillator strengths are ultimately needed to correctly interpret the spectra of neutron stars and magnetized white dwarfs. However, there is an additional complication that in order to do so, it also becomes necessary to account for the effect of strong electric fields which are also present in the atmospheres of these objects. Therefore, the energies and wave functions obtained herein are only part of the solution, and there is yet work to be done before this goal can be achieved. Moreover, as the magnetic field strength increases in the intense magnetic field regime, effects due to finite nuclear mass become increasingly relevant. In the current study, the mass of the nucleus is assumed to be infinite, and as such we have not carried out a suitable correction. One way to account for finite nuclear mass effects is to employ a scaling relationship wherein the energies determined at a certain magnetic field strength $\beta_Z$ for an infinite nuclear mass, would be related to the corresponding binding energies for a finite nuclear mass at a different value of the magnetic field strength $\tilde{\beta}_Z$ \cite{Schmelcher1999}. Such a correction becomes increasingly important at that upper end of magnetic field strengths investigated in this study. However, while accounting for this correction is important, it was not possible to do so since the errors of the pseudospectral method at such field strengths was $\approx 10^{-4} E_{Z,\infty}$ which is typically greater than the magnitude of the correction \cite{Schmelcher1999}. Additionally, the software developed here could be extended to tackle atoms with greater number of electrons, such as carbon or oxygen. It can also be extended towards a 2D post-HF framework such as CI of MCHF which would no doubt yield more accurate results. 
In either case, the method developed herein would be central to such enhancements and as such, the current study represents the very first implementation of a cylindrical pseudospectral method for atomic structure calculations in intense magnetic fields.

\appendix*
\section{\label{sec:pseudo} The Pseudo-spectral Approach}

For the numerical solution of the HF equations we employ a discretization based on pseudospectral methods extending our earlier investigation \cite{HT2010} to intense magnetic fields. However in contrast to our earlier work, in the current study we employ a cylindrical coordinate system. As a result the methodology for setting up the problem is considerably different from that described in Ref.~\cite{HT2010}. This section is arranged as follows. First we describe the methodology employed for solving the hydrogen atom using pseudospectral methods in cylindrical coordinates. Particular emphasis is placed on the implementation of boundary conditions. Thereafter, the treatment is extended to the particular case of the helium atom in a single configuration and a generalization of the scheme is then provided for multi-electron atoms. In the sections that follow, a considerable amount of detail is provided for the benefit of readers not familiar with pseudospectral methods, others may wish to only read Section~\ref{sec:helium} of the appendix.

\subsection{\label{sec:hydrogen}The Hydrogenic Problem}

We begin with the Hamiltonian for the hydrogenic problem (single-electron) in a strong magnetic field,
\begin{eqnarray}
\left[-\left(
\frac{1}{\rho_{i}}\frac{\partial}{\partial\rho_{i}}\left(\rho_{i}\frac{\partial}{\partial\rho_{i}}\right)+\frac{\partial^{2}}{\partial z_{i}^{2}}\right)
+\frac{({m_{i}})^{2}}{\rho_{i}^{2}} \right.\nonumber\\
\left.+\beta_{Z}^{2}\rho_{i}^{2}-\frac{2}{\sqrt{\rho_{i}^{2}+z_{i}^{2}}}\right]\psi_{i}\left(\rho_{i},z_{i}\right) = \epsilon_{i}\psi_{i}\left(\rho_{i},z_{i}\right).
\label{eq:7}
\end{eqnarray}
The solution of the eigenvalue problem in Eq.~(\ref{eq:7}) yields the individual electron energies and wave functions in a given configuration. The domains of both the radial and the axial coordinates are $0 \leqslant \rho, z < \infty$. The problem maintains azimuthal symmetry and thus, a solution of Eq.~(\ref{eq:7}) in this domain, when reflected about the $z=0$ plane (respecting $z-$parity of course) and revolved about the z-axis through $2\pi$, gives the solution in three-dimensional cylindrical coordinates. 

With the transformations given in Section~\ref{sec:numer} in Eqs.~(\ref{eq:transform_rho}) and (\ref{eq:transform_z}), we can re-write Eq.~(\ref{eq:7}) as
\begin{widetext}
\begin{eqnarray}
\left[ - \tilde{\alpha}^2 \frac{10^{-(x+1)} - 10^{-2(x+1)}}{10^{x+1}-1}  \frac{\partial^2}{\partial x^2}
-\tilde{\alpha}\alpha \frac{10^{-2(x+1)}}{10^{x+1}-1}  \frac{\partial}{\partial x} 
-\tilde{\alpha}^2  10^{-2(y+1)} \frac{\partial^2}{\partial y^2}
+ \tilde{\alpha}\alpha10^{-2(y+1)} \frac{\partial}{\partial y} + \right. \nonumber\\
\left. m_i^2 \left(\frac{\alpha}{10^{x+1}-1}\right)^2 
+ \beta_Z^2 \left(\frac{10^{x+1}-1}{\alpha} \right)^2
-\frac{2 \alpha}{\sqrt{(10^{x+1}-1)^2+(10^{y+1}-1)^2}} \right] \psi_{i}\left(x,y\right) = \epsilon_i \psi_{i}\left(x,y\right),
\label{eq:10}
\end{eqnarray}
\end{widetext}
where, we have dropped the subscripts on the coordinate labels. For brevity we have introduced the constant $\tilde{\alpha} = \alpha / \textrm{ln}(10) $. The discretisation points are thereafter taken to be the commonly used Chebyshev-Lobatto points \cite{HT2010,Trefethen,Boyd} given by
\begin{equation}
x_j = \textrm{cos}\left( \pi j / N\right),
\label{eq:11}
\end{equation}
where $j=0,1,...,N$. As is customary, we employ monic polynomials of degree $N$ as the cardinal functions to interpolate between these points and are given by \cite{HT2010,Trefethen}, 
\begin{equation}
p_j(x) = \frac{1}{a_j}\prod_{\substack{
					k=0\\
					k \neq j}}^{N} (x-x_k),
\label{eq:12}
\end{equation}
with 
\begin{equation}
a_j = \prod_{\substack{
					k=0\\
					k \neq j}}^{N} (x_j-x_k).
\label{eq:13}
\end{equation}
Derivatives of these interpolating polynomials at the discretisation points then yield the so-called Chebyshev differntiation matrix, whose elements are given by, 
\begin{equation}
D_{ij} = \frac{1}{a_j} \prod_{\substack{
					k=0\\
					k \neq i,j}}^{N} (x_i-x_k) = \frac{a_i}{a_j (x_i - x_k)}~~(i \neq j)
\label{eq:14}
\end{equation}
and
\begin{equation}
D_{jj} =\sum_{\substack{
					k=0\\
					k \neq j}}^{N} (x_j - x_k)^{-1}
\label{eq:15}
\end{equation}
In writing Eq.~(\ref{eq:10}), we have removed the co-ordinate singularity at $x=-1$, by replacing it with $x=-1+\delta$, where $\delta = 10^{-14}$ in units of Bohr radii. This approximation produced acceptable results within error tolerances. The outer boundary conditions of the domain, at $x=1~\textrm{and}~y=1$ (corresponding to $\rho, z =\infty$), are taken care of by imposing Dirichlet boundary conditions since the wave function must vanish at infinity (see below). The remaining inner boundaries of the compactified domain at $x,y= - 1$ can have either Dirichlet or Neumann boundary conditions, depending upon the wave function in question. The following discussion delineates the methodology for the 2D problem.

\subsubsection{\label{sec:discrete}An Explicit Example - Domain Discretization}

We consider here an explicit example to illustrate the use of pseudospectral methods for solving an eigenvalue problem. The method developed here is a non-trivial extension of the one developed by the authors in Ref.~\cite{Talbot2005}. 

\begin{figure}[h]
\begin{center}
\includegraphics[width=3.5in, scale=1.0]{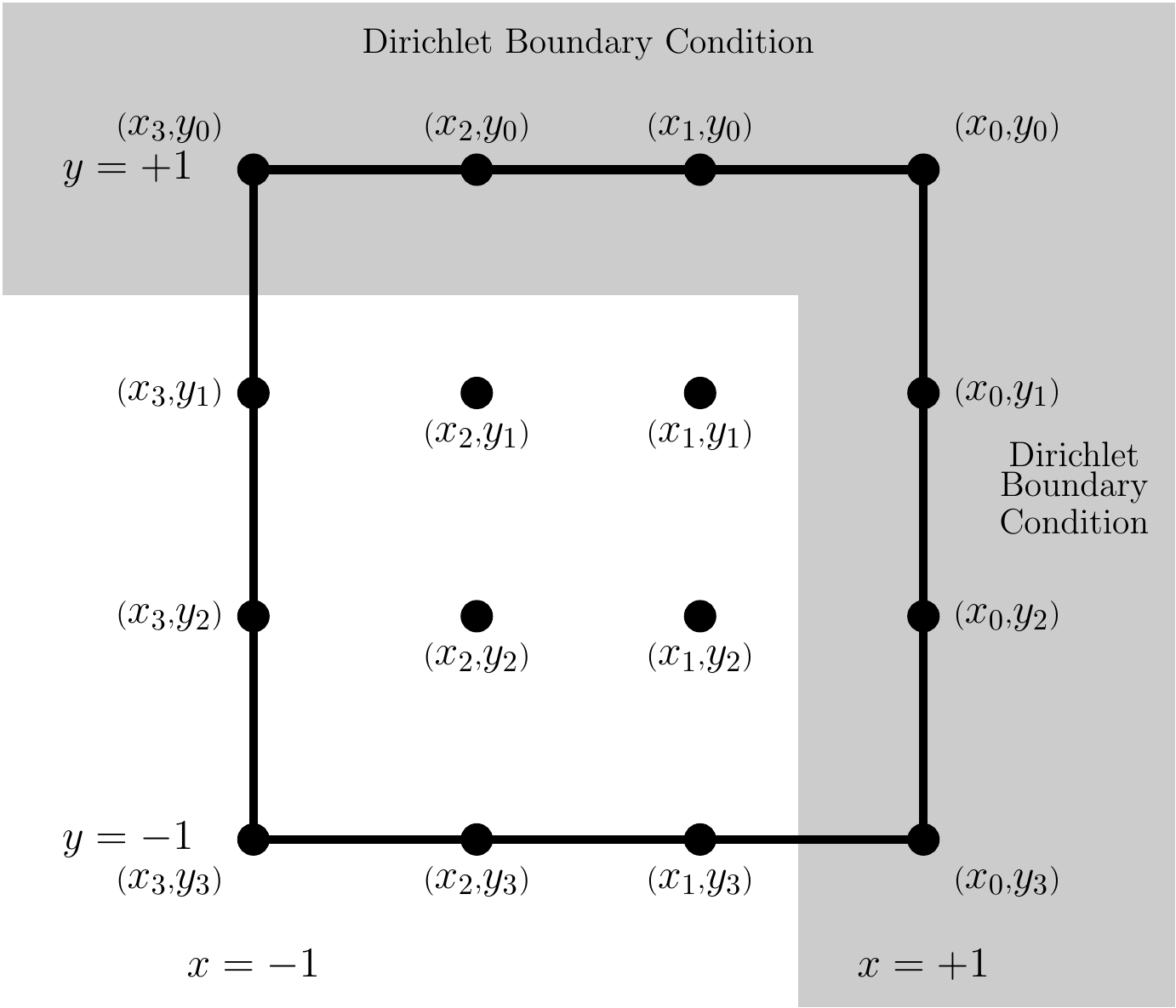}
\end{center}
\caption{Pictorial representation of the domain $[-1,1] \otimes [-1,1]$ discretised using $N+1$ points in each direction, with $N=3$. The outer boundaries have Dirichlet conditions imposed. } 
\label{fig:fig1}
\end{figure}

Let us begin with a domain $[-1,1] \otimes [-1,1]$ which is discretised using $N+1$ points in each of the two Cartesian directions; $x$ and $y$, with $N=3$ in this explicit example. This is illustrated in Fig.~(\ref{fig:fig1})
The partial differential equation in Eq.~(\ref{eq:10}) is two-dimensional therefore, following the procedure outlined in Refs.~\cite{HT2010,Talbot2005}, we can construct two-dimensional operators by employing Kronecker products of matrices, for example we can define, 
\begin{equation}
\tilde{D}_x^2 = \left[D_x \times D_x \right] \otimes I_y
\label{eq:31}
\end{equation}
and
\begin{equation}
\tilde{D}_y^2 = I_x \otimes \left[D_y \times D_y \right] ,
\label{eq:32}
\end{equation}
where $D_x$ and $D_y$ are Chebyshev differentiation matrices in the $x-$ and $y-$ directions respectively, defined similar to Eqs.~\ref{eq:14} and \ref{eq:15}. The matrices $I_x$ and $I_y$ are identity matrices of dimension $N_x \times N_x$ and $N_y \times N_y$ respectively. The matrices $\tilde{D}_x^2$ and $\tilde{D}_y^2$ are discrete representations of the $\partial^2 / \partial x^2$ and $\partial^2 / \partial y^2$ operators, respectively. In these matrices the entries corresponding to the outer boundaries at $x,y=+1$ have already been excised due to Dirichlet boundary conditions \cite[][see Figure~\ref{fig:fig1}]{HT2010,Trefethen}. As a result, the indices $i,j$ of these matrices are limited to an upper value of $N$ rather than $N+1$. In our example, the number of node points in the $x$ and $y$-directions are equal, therefore $N_x=N_y=N$. Thus, using these Chebyshev differentiation matrices we can write down Eq.~(\ref{eq:10}) in matrix form as,
\begin{align}
\left[ \textrm{diag}(a) \times \tilde{D}_{x}^2 + \textrm{diag}(b) \times \tilde{D}_x + \textrm{diag}(c) \times \tilde{D}_y^2 \right.\nonumber\\ 
\left. + \textrm{diag}(d) \times \tilde{D}_y + \textrm{diag}(e) \right] \psi_{i} \equiv L_i \psi_i = \epsilon_i \psi_{i},
\label{eq:33}
\end{align}
where $\tilde{D}_x = D_x \otimes I_y$ and $\tilde{D}_y = I_x \otimes D_y$, with Dirichlet boundary conditions imposed on the outer boundaries and therefore once again appropriately trimmed. 
The diagonal matrices $a,b,c,d$ and $e$ are the coefficients of the different terms in Eq.~(\ref{eq:10}).
Replacing $\psi_i$ with the polynomials in Eq.~(\ref{eq:12}), the collocation points of the problem then forms a mesh with the corresponding values $p(x_i,y_j)$ with $i,j=0,...,N$. The collocation points are those illustrated in Fig.~(\ref{fig:fig1}). However, instead of writing the polynomial as a matrix of values at the collocation points, we can write the matrix as an extended vector comprising of the different columns, one followed by another. This then forms an $N^2 \times 1$ vector rather than an $N \times N$ matrix. Explicitly, we can reshape the matrix $p(x_i,y_j)$, with values given at the collocation points to form,
\begin{equation}
\textbf{p}=\begin{bmatrix} 
				p(x_1,y_1) \\
				p(x_1,y_2) \\
				p(x_1,y_3) \\
				p(x_2,y_1) \\
				p(x_2,y_2) \\
				p(x_2,y_3) \\
				p(x_3,y_1) \\
				p(x_3,y_2) \\
				p(x_3,y_3) \\
		\end{bmatrix}.
\label{eq:36}			
\end{equation}		
It is to be remembered that in our explicit example $N=3$. 

\subsubsection{\label{sec:bc}Boundary Condition Implementation}

\begin{figure}[h]
\begin{center}
\includegraphics[width=3.5in, scale=1.0]{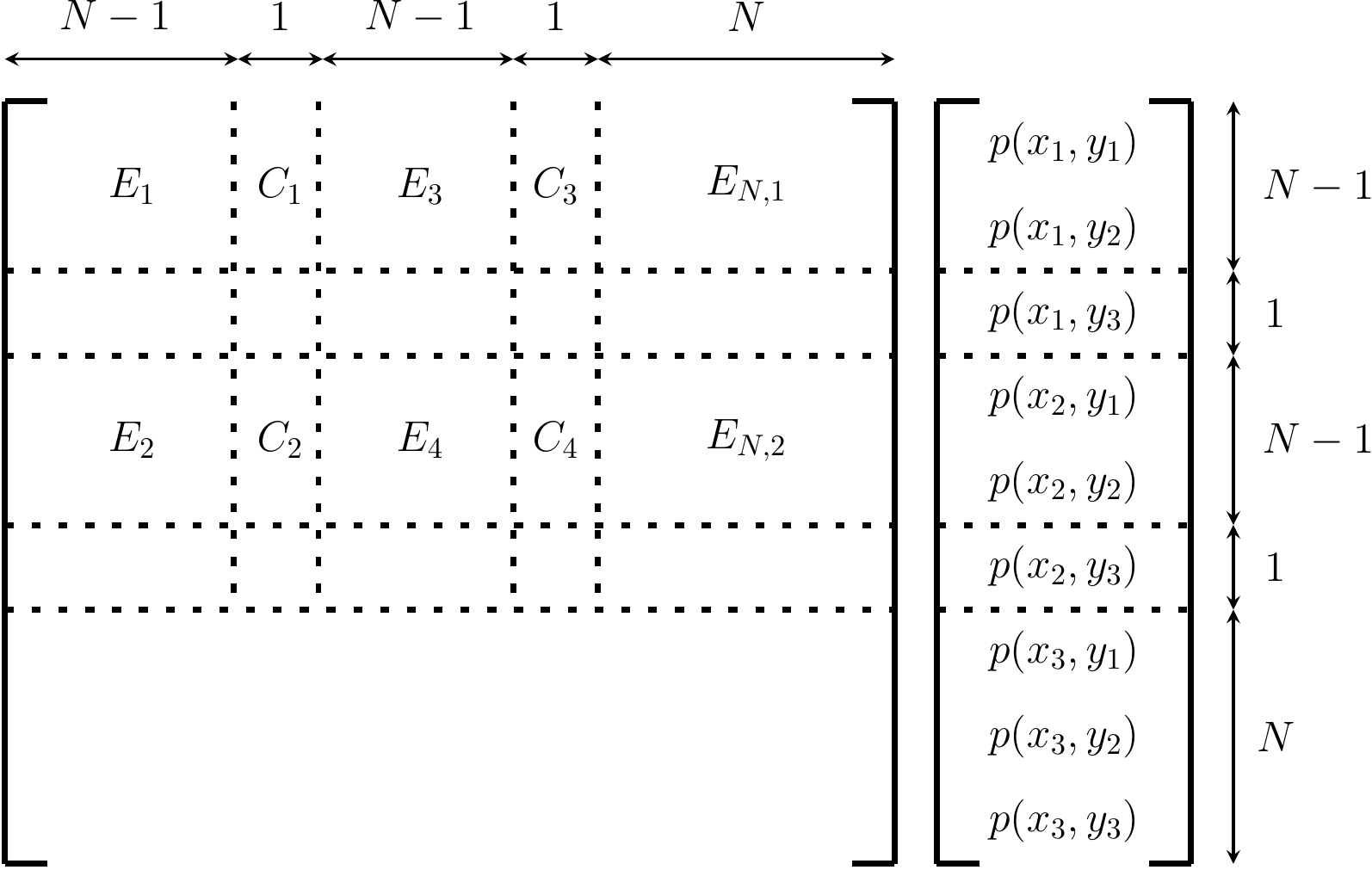}
\end{center}
\caption{Pictorial representation the operator $L$ acting upon the vector $\textbf{p}$. The number of points in either direction, is $N=3$ in the current example.} 
\label{fig:fig2}
\end{figure}

\begin{figure}[h]
\begin{center}
\includegraphics[width=3.5in, scale=1.0]{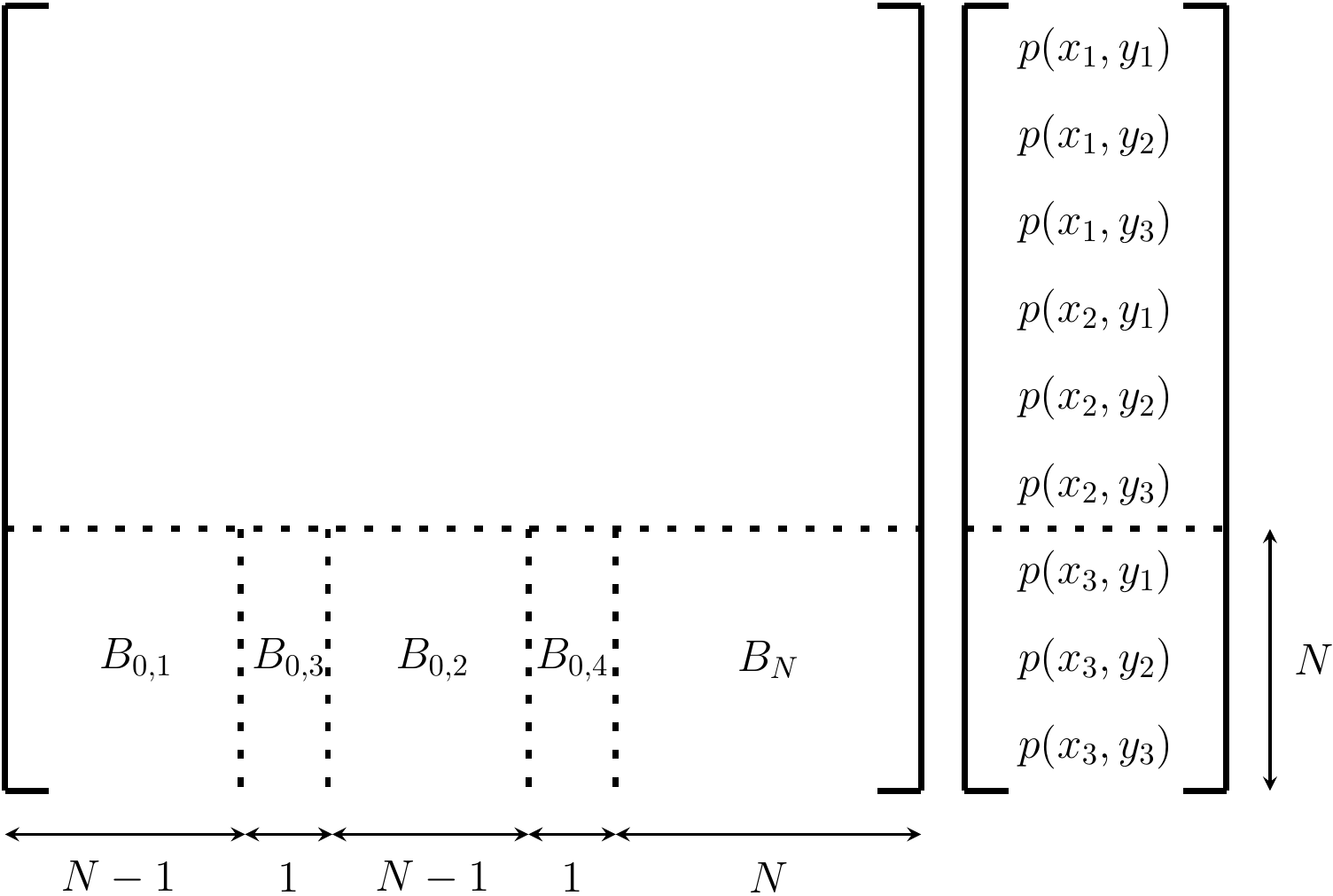}
\end{center}
\caption{Pictorial representation the operator $B_x$ acting upon the vector $\textbf{p}$. The number of points in either direction, is $N=3$.} 
\label{fig:fig3}
\end{figure}

\begin{figure}[h]
\begin{center}
\includegraphics[width=3.5in, scale=1.0]{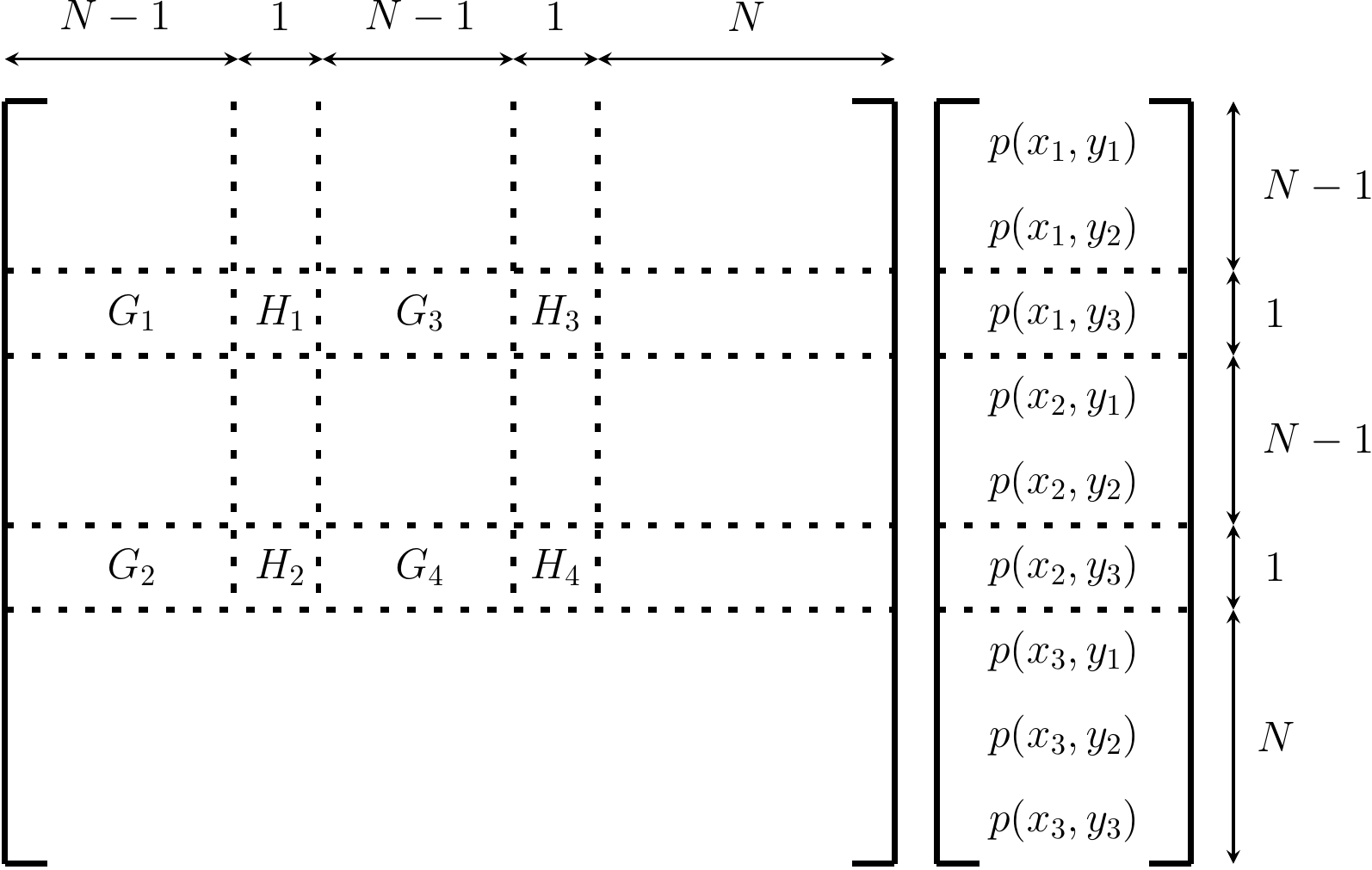}
\end{center}
\caption{Pictorial representation the operator $B_y$ acting upon the vector $\textbf{p}$. The number of points in either direction, is $N=3$.} 
\label{fig:fig4}
\end{figure}

Presently, Eq.~(\ref{eq:33}) can be re-cast into matrix form using the interpolating polynomials as,
\begin{equation}
L\textbf{p}=\lambda \textbf{p},
\label{eq:37}
\end{equation}
where the eigenvalues of the spectrum $\lambda$ are the
individual single particle energies. Let us now
suppose that we wish to solve the eigenvalue problem for the $1s_0$
state of the hydrogen atom.  Using the notation for labeling the states in the presence of a magnetic field, the state $1s_0$ of hydrogen in the presence of a strong field would be written as $1^{1}0^{+}$. For this state of hydrogen the boundary conditions are as follows. Along both the $x,y = +1$ boundaries the wave function must vanish, therefore we have Dirichlet boundary conditions. Along the $x,y = -1$ boundaries however, we have Neumann conditions. These boundary conditions are to be kept in mind for the following discussion. A pictorial representation of the operator $L$ acting upon the vector $\textbf{p}$ is given in Fig.~(\ref{fig:fig2}),
which shows the portions of the matrix operator $L$ that are relevant to the solution in the interior mesh points of the problem. These are the sub-matrices $E_1,E_2,E_3,E_4$, $C_1,C_2,C_3,C_4$,$E_{N,1}$ and $E_{N,2}$.
The vector $\textbf{p}$ can accordingly be split into three components,
\begin{subequations}
\begin{equation}
\textbf{p}_\textrm{int}=\begin{bmatrix} 
				p(x_1,y_1) \\
				p(x_1,y_2) \\
				p(x_2,y_1) \\
				p(x_2,y_2) \\
				p(x_2,y_3) \\
		\end{bmatrix},
\label{eq:38a}	
\end{equation}
\begin{equation}
\textbf{p}_{b_y}=\begin{bmatrix} 
				p(x_1,y_3) \\
				p(x_2,y_3) \\
		\end{bmatrix}
\label{eq:38b}	
\end{equation}
and
\begin{equation}
\textbf{p}_{b_x}=\begin{bmatrix} 
				p(x_3,y_1) \\
				p(x_3,y_2) \\
				p(x_3,y_3) \\				
		\end{bmatrix}.
\label{eq:38c}	
\end{equation}
\end{subequations}
In the above, $\textbf{p}_\textrm{int}$ refers to the function value in the interior points of the mesh given in Fig.~(\ref{fig:fig1}), while the function values at the $x=-1$ boundary are given by $\textbf{p}_{b_x}$ and correspondingly, the values at the $y=-1$ boundary are given by $\textbf{p}_{b_y}$. The eigenvalue problem in the interior mesh points is then given by,
\begin{equation}
E \textbf{p}_\textrm{int} = \lambda \textbf{p}_\textrm{int} - C \textbf{p}_{b_y} - E_{N} \textbf{p}_{b_x}.
\label{eq:39}
\end{equation}
The matrices $E$, $C$ and $E_N$ are given by (see Fig.~(\ref{fig:fig2})),
\begin{subequations}
\begin{equation}
E=\begin{bmatrix} 
		E_1~~E_3 \\
		E_2~~E_4 \\
    \end{bmatrix}_{(N-1)^2 \times (N-1)^2},
\label{eq:40a}
\end{equation}
\begin{equation}
C=\begin{bmatrix} 
		C_1~~C_3 \\
		C_2~~C_4 \\
    \end{bmatrix}_{(N-1)^2 \times (N-1)}
\label{eq:40b}
\end{equation}
and
\begin{equation}
E_N=\begin{bmatrix} 
		E_{N,1} \\
		E_{N,2} \\
    \end{bmatrix}_{(N-1)^2 \times N}.
\label{eq:40c}
\end{equation}
\end{subequations}
The function values at the boundaries, i.e., $\textbf{p}_{b_x}$ and $\textbf{p}_{b_y}$ are unknown and can be expressed in terms of the solution in the interior $\textbf{p}_\textrm{int}$ using the boundary conditions. A Neumann boundary condition imposed on the wave function $\psi$ along a given boundary ($\partial \Omega$) is given by,
\begin{equation}
\hat{\textbf{n}} \cdot \bf{\nabla} \psi |_{\partial \Omega} = g,
\label{eq:41}
\end{equation}
where $\hat{\textbf{n}}$ is the unit vector normal to the boundary and $g$ is the value to which the directional derivative of the function along the direction of the normal vector is set. In our case, the boundaries in question are the lines $x,y=-1$. The corresponding normal vectors are then trivial and we obtain the conditions (using Chebyshev differentiation matrices),
\begin{equation}
\left.B_x \textbf{p} \right|_{x=-1} = 0
\label{eq:44}
\end{equation}
and 
\begin{equation}
\left.B_y \textbf{p} \right|_{y=-1} = 0.
\label{eq:45}
\end{equation}
The boundary matrices are defined by $B_x=D_x \otimes I_y$ and $B_y = I_x \otimes D_y$. By virtue of the Kronecker product with the identity matrix, $B_x$ has entries only along the diagonals. The matrix $B_y$ on the other hand, is a block diagonal matrix with each block of dimension $N \times N$.
Fig.~(\ref{fig:fig3}) shows a pictorial representation of the boundary matrix $B_x$ acting on the vector $\textbf{p}$.
In this case, since the derivative is required to vanish along the $x=-1$ boundary, we focus our attention only on the part of the boundary matrix $B_x$, that acts on the vector $\textbf{p}$ along the boundary in question, i.e., the last $N$ rows. Fig.~(\ref{fig:fig3}) shows the sub-matrices that are needed, these are labelled $B_{0,1},~B_{0,2},~B_{0,3},~B_{0,4}$ and $B_N$. Thus, we can write the relevant portion of Eq.~(\ref{eq:44}) as,
\begin{equation}
B_1 \textbf{p}_\textrm{int} + B_2 \textbf{p}_{b_y} + B_N \textbf{p}_{b_x} = 0.
\label{eq:46}
\end{equation}
The matrices $B_1$ and $B_2$ in Eq.~(\ref{eq:46}) are given by,
\begin{subequations}
\begin{equation}
B_1= \begin{bmatrix} 
		B_{0,1} ~~ B_{0, 2} \\
         \end{bmatrix}_{N \times (N-1)^2}
\label{eq:47a}
\end{equation}
and
\begin{equation}
B_2=\begin{bmatrix} 
		B_{0,3} ~~ B_{0,4} \\
        \end{bmatrix}_{N \times (N-1)}
\label{eq:47b}
\end{equation}
\end{subequations}
The dimensions of the matrices are indicated for convenience. Eq.~(\ref{eq:46}) has two unknowns, $\textbf{p}_{b_x}$ and $\textbf{p}_{b_y}$ that need to be expressed in terms of $\textbf{p}_\textrm{int}$. Therefore we need another equation. This is provided by Eq.~(\ref{eq:45}). We have shown in Fig.~(\ref{fig:fig4}) a pictorial representation of the boundary matrix $B_y$ acting upon the the vector $\textbf{p}$. Once again we only focus our attention on the part of $B_y$ that acts on the inner $y-$boundary.
The different sub-matrices are also shown in Fig.~(\ref{fig:fig4}). We can then write down an equation for expressing the boundary condition in Eq.~(\ref{eq:45}) as,
\begin{equation}
G \textbf{p}_\textrm{int} + H \textbf{p}_{b_y} = 0.
\label{eq:48}
\end{equation}
The matrices $G$ and $H$ are given by,
\begin{subequations}
\begin{equation}
G = \begin{bmatrix} 
		G_1 ~~ G_3 \\
		G_2 ~~ G_4 \\
         \end{bmatrix}_{(N-1) \times (N-1)^2}
\label{eq:49a}
\end{equation}
and
\begin{equation}
H  =\begin{bmatrix} 
		H_1 ~~ H_3 \\
		H_2 ~~ H_4 \\
         \end{bmatrix}_{(N-1) \times (N-1)}
\label{eq:49b}
\end{equation}
\end{subequations}
Since $B_y$ is a block diagonal matrix, with each block of dimension $N \times N$, only the block diagonal portion of $G$ and the diagonal of $H$, are respectively non-zero. In the current explicit example, we therefore have $G_2 = G_3 = [~0 ~~ 0~] $ and $H_2 = H_3 = 0$. With these definitions in place, we can employ Eqs.~(\ref{eq:46}) and (\ref{eq:48}) to obtain expressions for the unknowns $\textbf{p}_{b_x}$ and $\textbf{p}_{b_y}$ in terms of the values in the interior $\textbf{p}_\textrm{int}$ as,
\begin{equation}
\textbf{p}_{b_y} = - H^{-1}G \textbf{p}_\textrm{int}
\label{eq:50}
\end{equation}
and
\begin{equation}
\textbf{p}_{b_x} = -B_N^{-1}(B_1 - B_2 H^{-1}G) \textbf{p}_\textrm{int}.
\label{eq:51}
\end{equation}
Substituting Eqs.~(\ref{eq:50}) and (\ref{eq:51}) in Eq.~(\ref{eq:39}) we obtain an eigenvalue problem for the solution in the interior mesh points as,
\begin{eqnarray}
\left(E - E_N B_N^{-1} (B_1 - B_2 H^{-1}G) ~ - \right. \nonumber\\
\left.C H^{-1} G\right) \textbf{p}_\textrm{int} =  \lambda \textbf{p}_\textrm{int}.
\label{eq:52}
\end{eqnarray}
Eq.~(\ref{eq:52}) can then be solved as an eigenvalue problem using standard methods to obtain the eigenvalues $\lambda$ and the eigenvectors. See Section \ref{sec:numer} for details regarding the numerical methods employed. For the moment, we turn our attention to extending this methodology to the two-electron problem. 

\subsection{\label{sec:helium}The Two-Electron Problem}

The HF problem for the two electron atom can be written, using the matrix formalism detailed above in a compact form as,
\begin{equation}
	\begin{pmatrix}
		L_1  + \frac{2}{Z} \textrm{diag}[\Phi_{D,1}] ~~ - \frac{2}{Z} \textrm{diag}[\alpha_E] \\
		 \\
		- \frac{2}{Z} \textrm{diag}[\alpha_E] ~~ L_2  + \frac{2}{Z} \textrm{diag}[\Phi_{D,2}]  \\		
	\end{pmatrix}
	\begin{pmatrix}
		\psi_1 \\
		\\
		\psi_2 \\
	\end{pmatrix}
	=
	\lambda
	\begin{pmatrix}
		\psi_1 \\
		\\
		\psi_2 \\
	\end{pmatrix}.	
\label{eq:53}
\end{equation}
It is evident upon inspection that Eq.~(\ref{eq:53}) is a coupled eigenvalue problem. The operators $L_1$ and $L_2$ are the operators defined in Eq.~(\ref{eq:33}). The direct and exchange operators are $\textrm{diag}[\Phi_{D,i}]$ and $\textrm{diag}[\alpha_{E}]$ respectively, and the latter makes the problem non-linear, as it depends upon the solutions $\psi_i$. However, the problem is linearized by estimating the direct and exchange interactions using wave functions from the previous iteration. The exchange interaction still couples the two electrons and as such, we are still required to solve a coupled eigensystem.

To continue with our explicit example, let us suppose that we wish to calculate the energy of the helium atom in the configuration $1^3(-1)^+$, or in terms of field-free notation, $1s_02p_{-1}$. Thus the hydrogenic problem would first need to be solved for each of the two electrons in the configuration. Let us label the electrons' wave functions using $\textbf{p} \equiv 1s_0$ and $\textbf{q} \equiv 2p_{-1}$. 
Presently, we briefly describe the method of solution of the elliptic partial differential equations for obtaining the direct and exchange interactions.

\subsubsection{\label{sec:direx}The Direct and Exchange Interactions}

Let us assume that we have solved the hydrogenic problem and already
obtained initial estimates for the wave functions of each of the two
electrons, viz., $\textbf{p}$ and $\textbf{q}$ for the states $1s_0$
and $2p_{-1}$, using the method described in
Section~\ref{sec:hydrogen}. Eqs.~(\ref{eq:3}) and (\ref{eq:4}) in Section~\ref{sec:HF} can be re-written after domain compactification in a compact form employing the matrix formulation presented above as,
\begin{equation}
L_\textrm{dir} \Phi_{D,i} = -4 \pi \left \{ \begin{array}{cl}
\textbf{p}^2 & \textrm{for}~i=2 \\
\textbf{q}^2 & \textrm{for}~i=1 ,
  \end{array}
\right .
\label{eq:56}
\end{equation}
for the direct interactions, and as
\begin{equation}
L_\textrm{exch} \alpha_{E} = -4 \pi (\textbf{p} \textbf{q}),
\label{eq:57}
\end{equation}
for the exchange interaction between the electrons.
In the above, $i=1 ~\textrm{or}~ 2$, labels the electrons, for the two-electron problem. 
$L_\textrm{dir}$ and $L_\textrm{exch}$ are the left hand side operators in Eqs.~(\ref{eq:3}) and (\ref{eq:4}) respectively (see Section~\ref{sec:HF}).

The boundary conditions for the direct and exchange interactions actually result in a rather simple implementation. If the interaction between two wave-functions has $\Delta m  \neq 0$ then the interaction potential must vanish along the magnetic axis \cite{TH2009, HT2010}. Therefore for the direct interactions, since $\Delta m = 0$ for the electron's interaction with itself, the interaction potential does not vanish along the magnetic axis. On the other hand, for the exchange interaction between $\textbf{p} \equiv 1s_0$ and $\textbf{q} \equiv 2p_{-1}$, the interaction potential must vanish along the magnetic axis since $\Delta m = -1$. These collectively imply that we need to set up Neumann boundary conditions along both $x=-1$ and $y=-1$, for the direct interactions. Whereas for the exchange interactions we set up Neumann conditions along $y=-1$, but a Dirichlet boundary condition along $x=-1$ if $\Delta m \neq 0$, which is the case in our explicit example for the $1s_0$ and $2p_{-1}$ electrons.



With these boundary conditions now identified, they can be imposed on the operators $L_\textrm{dir}$ and $L_\textrm{exch}$ using the methods described in Section~\ref{sec:bc}. The two linear systems of equations at the collocation points, given by Eqs.~(\ref{eq:56}) and (\ref{eq:57}), are solved using standard methods, to obtain the direct and exchange potentials as,
\begin{subequations}
\begin{align}
\Phi_{D,1}=\left\{E_\textrm{dir} - E_{N,\textrm{dir}}~ B_N^{-1} (B_1 - B_2 H^{-1}G) ~ - \right. \nonumber\\
\left. C H^{-1} G\right\}^{-1} \textbf{q}_\textrm{int}^2.
\label{eq:58a}
\end{align}
\begin{align}
\Phi_{D,2}=\left\{E_\textrm{dir} - E_{N,\textrm{dir}}~ B_N^{-1} (B_1 - B_2 H^{-1}G) ~ - \right. \nonumber\\
\left. C H^{-1} G\right\}^{-1} \textbf{p}_\textrm{int}^2.
\label{eq:58b}
\end{align}
and
\begin{eqnarray}
\alpha_{E}=\left\{E_\textrm{exch} ~ - C H^{-1} G\right\}^{-1} (\textbf{pq})_\textrm{int} .
\label{eq:58c}
\end{eqnarray}
\end{subequations}
In the above, the matrices $E_\textrm{dir}$, $E_{N,dir}$ and $E_\textrm{exch}$ are defined similarly to Eqs.~(\ref{eq:40a}) and (c), this time however, using the operators $L_\textrm{dir}$ and $L_\textrm{exch}$ respectively (see Section~\ref{sec:bc}). Once the direct and exchange interactions have been determined, they can be substituted in Eq.~(\ref{eq:53}) and the coupled eigensystem can be solved.

Note that the boundary conditions implemented in the explicit example are specifically for the configuration of the helium atom given by $1^3(-1)^+$ or $1s_02p_{-1}$. For other configurations, the boundary conditions imposed on $\textbf{p}$ and $\textbf{q}$ would be different. In that case, Eqs.~(\ref{eq:58a}-c) would change accordingly. With this in mind, we now proceed to the next section which describes the setup of the coupled eigenvalue problem in Eq.~(\ref{eq:53}) and the implementation of boundary conditions for its solution.

\subsubsection{\label{sec:coupled_eig}The Coupled Eigenvalue Problem}

The direct and exchange potentials are found in Eqs.~(\ref{eq:58a}-c) as vectors. These are converted to matrices with entries on the main diagonal before substituting into Eq.~(\ref{eq:53}). If we label the operator on the left hand side of Eq.~(\ref{eq:53}) as $M$, then we can re-write Eq.~(\ref{eq:53}) as,
\begin{equation}
\begin{pmatrix}
	M_{11} ~~ M_{12} \\
	\\
	M_{21} ~~ M_{22} \\
\end{pmatrix}
\begin{pmatrix}
	\textbf{p} \\
	\\
	\textbf{q} \\
\end{pmatrix}
=
\lambda
\begin{pmatrix}
	\textbf{p} \\
	\\
	\textbf{q} \\
\end{pmatrix}.
\label{eq:59}
\end{equation} 

We can depict pictorially, the action of $M$ on the vector $\begin{pmatrix} \psi_1 \\ \psi_2 \\ \end{pmatrix}$ or equivalently $\begin{pmatrix} \textbf{p}\\ \textbf{q}\\ \end{pmatrix}$, as shown in Fig.~(\ref{fig:fig5}). In the matrices $M_{11}$ and $M_{22}$, the off-diagonal sub-matrices are identical to those in Fig.~(\ref{fig:fig3}), see Eqs.~(\ref{eq:40a}-c). Also, it can be seen that the matrices $M_{12}$ and $M_{21}$ are diagonal matrices that are identical. Only the non-zero parts of these matrices that act on the interior parts of the vectors, $\textbf{p}_\textrm{int}$ and $\textbf{q}_\textrm{int}$ are shown in Fig.~(\ref{fig:fig5}). A Dirichlet boundary condition has been imposed explicitly along $x=-1$ for the vector $\textbf{q}$ by setting $\textbf{q}_{b_x}=0$, as shown in Fig.~(\ref{fig:fig5}). The coupled eigensystem can then be written as a system of coupled matrix equations for the interior points as,
\begin{subequations}
\begin{eqnarray}
E^{11} \textbf{p}_\textrm{int} + T^{12} \textbf{q}_\textrm{int} = \lambda \textbf{p}_\textrm{int} - C \textbf{p}_{b_y} - E_N \textbf{p}_{b_x}
\label{eq:60a}
\end{eqnarray}
and
\begin{eqnarray}
T^{12} \textbf{p}_\textrm{int} + E^{22} \textbf{q}_\textrm{int} = \lambda \textbf{q}_\textrm{int} - C \textbf{q}_{b_y} .
\label{eq:60b}
\end{eqnarray}
\end{subequations}
The sub-matrices $C$ and $E_N$ are defined as given in Eqs.~(\ref{eq:40b} \& c). However, The sub-matrix $E^{11}$ and $E^{22}$ have slightly different entries on the diagonal and are thus defined as,
\begin{equation}
E^{ii}=\begin{bmatrix} 
		E_1^{ii}~~E_3 \\
		E_2~~E_4^{ii} \\
    \end{bmatrix}_{(N-1)^2 \times (N-1)^2},~ i=1,2.
\label{eq:61}
\end{equation}
The matrix $T^{12}$ is a block diagonal matrix comprised of $T^{12}_1$ and $T^{12}_2$ as shown in Figure~\ref{fig:fig5}. Similar to our discussion regarding the hydrogen atom in Section~\ref{sec:bc}, we are required to express the vectors $\textbf{p}_{b_x}$, $\textbf{p}_{b_y}$ and $\textbf{q}_{b_y}$ in terms of $\textbf{p}_\textrm{int}$ and $\textbf{q}_\textrm{int}$ respectively, by implementing Neumann boundary conditions. This would enable us to then cast the coupled eigenvalue problem into its final form as,
%
\begin{widetext}
\begin{eqnarray}
\begin{pmatrix}
	\left[E^{11} - C H^{-1}G - E_N B_N^{-1}(B_1 - B_2 H^{-1}G)\right] ~~~~~~~~~~ T^{12} ~~~~~ \\
	\\
	~~~~~ ~~~~~ ~~~~~ ~~~~~ ~~~~~ ~~~~~T^{12} ~~~~~  ~~~~~ ~~~~~ ~~~~~ ~~~~~ ~~~~~ \left[E^{22} - C H^{-1}G\right] \\
\end{pmatrix}
\begin{pmatrix}
	\textbf{p}_\textrm{int} \\
	\\
	\textbf{q}_\textrm{int} \\
\end{pmatrix}
=
\lambda
\begin{pmatrix}
	\textbf{p}_\textrm{int} \\
	\\
	\textbf{q}_\textrm{int} \\
\end{pmatrix}.
\label{eq:64}
\end{eqnarray} 
\begin{figure}[h]
\begin{center}
\includegraphics[scale=1.0]{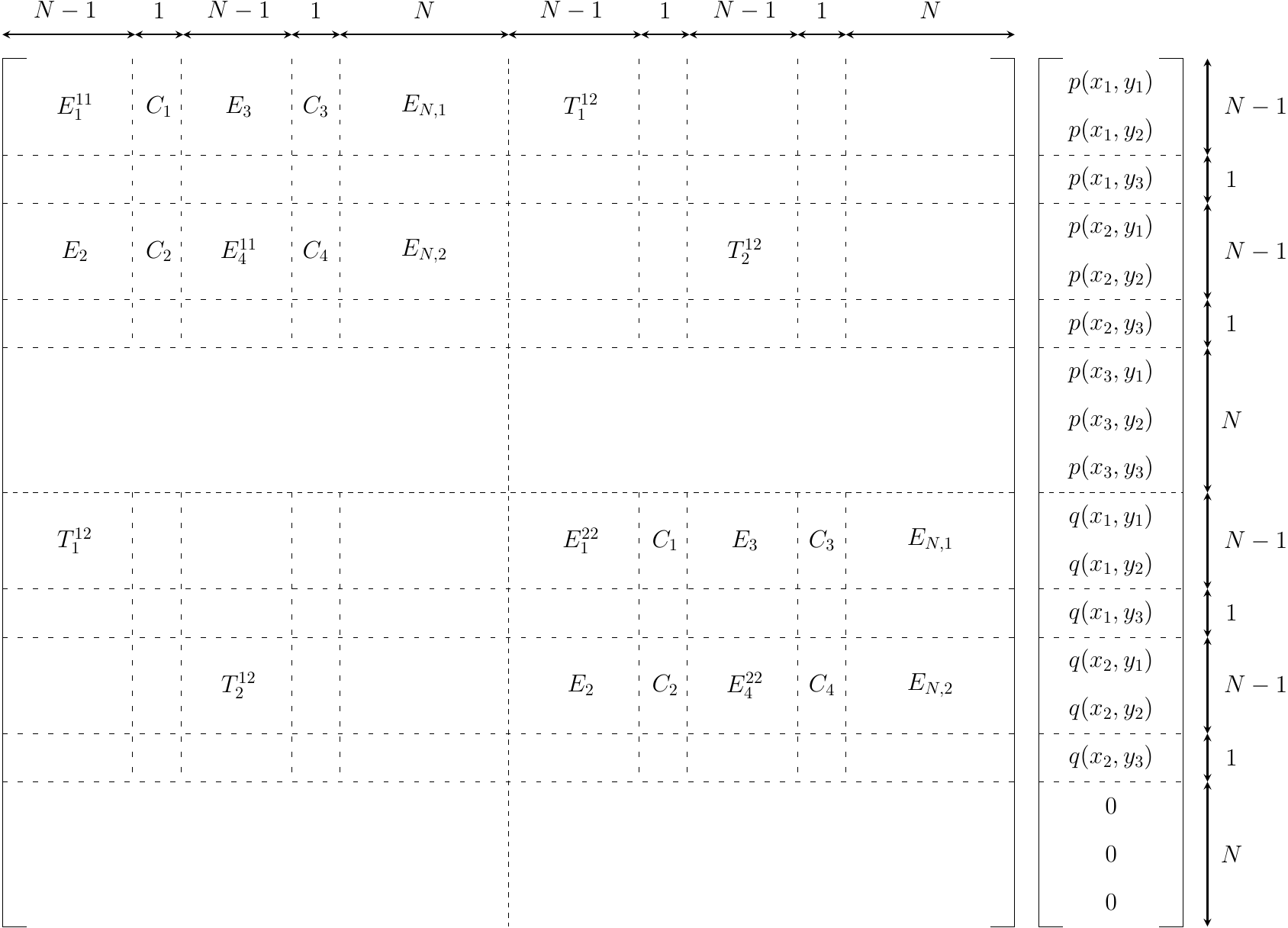}
\end{center}
\caption{Pictorial representation the operator $M$ acting upon the vector $\begin{pmatrix} \textbf{p}\\ \textbf{q}\\ \end{pmatrix}$. The number of points in either direction, is $N=3$.} 
\label{fig:fig5}
\end{figure}
\end{widetext}
The exchange operators are symmertric under permutation, thus $T^{ij} = T^{ji}, ~ i,j=1,2,3...$. The eigenvalue problem is uncoupled for each electron by expressing the exchange interactions as a potentials using wave functions from the previous iteration \cite{Slater1951}. With this Eq.~(\ref{eq:64}) is then solved using standard algorithms \cite{ARPACK} to obtain the eigenvectors and eigenvalues and the HF iterations are carried out until convergence. We would like to caution the reader once more that the formulation of the HF problem shown in Eq.~(\ref{eq:64}) is for the explicit example of the configuration of helium given by $1^3(-1)^+$ or $1s_02p_{-1}$. For other configurations, depending upon the boundary conditions, Eq.~(\ref{eq:64}) will take a very different form. The left hand side operator shown in Eq.~(\ref{eq:64}) is the pseudospectral representation of the HF operator for a particular configuration of the helium atom. 
The method described above can easily be extended to tackle the case of the few-electron atom, say lithium. It can be seen that the problem size in Eq.~(\ref{eq:64}) will grow not only with the number of mesh points but also with the number of electrons. For a given number of mesh points $N$ in each direction and a certain number of electrons $n_e$, the size of the pseudospectral HF operator in Eq.~(\ref{eq:64}) is $\left[n_e (N-1)^2) \times (n_e (N-1)^2\right]$. Using Slater's approximation for the exchange \cite{Slater1951} and uncoupling the eigenvalue problem, one would then solving $n_e$ individual eigenvalue problems each of size $\left[(N-1)^2 \times (N-1)^2\right]$. Thus, since computer memory requirements are governed by this latter system size, the coupled problem in Eq.~(\ref{eq:64}) is readily seen to be far more intensive than the uncoupled problem. 

\begin{acknowledgments}
This research was supported by funding from NSERC.  The calculations
were performed on computing infrastructure purchased with funds from
the Canadian Foundation for Innovation and the British Columbia
Knowledge Development Fund.
\end{acknowledgments}


\bibliography{PaperIII}


\end{document}